\documentclass[12pt]{article}

\pdfoutput=1

\usepackage{amsmath,amssymb,amsthm,mathrsfs,bbm}
\usepackage{latexsym,amscd,amsbsy,dsfont,amsfonts}
\usepackage{graphicx}

\usepackage{bm}
\usepackage{latexsym}
\usepackage[latin1]{inputenc}
\usepackage{amsfonts}
\usepackage{amssymb}
\usepackage{amsmath}
\usepackage{subfigure}
\usepackage{ccaption}    
\usepackage{tabularx}

\usepackage[
      colorlinks=true,
      linkcolor=blue,
      urlcolor=blue,
      filecolor=black,
      citecolor=red,
      pdfstartview=FitV,
      pdftitle={},
        pdfauthor={},
        pdfsubject={},
        pdfkeywords={},
        pdfpagemode={},
        bookmarksopen=true
      ]{hyperref}

\numberwithin{equation}{section}

\renewcommand{\Im}{\operatorname{Im}}
\renewcommand{\Re}{\operatorname{Re}}
\renewcommand{\ss}{s}

 \captionnamefont{\bfseries}

\title{\bf Relaxed superconductors}

\author{\large Tom\'as Andrade$^{a}$ and Simon A.~Gentle$^b$ \\ \\
	\small $^a$\it Rudolf Peierls Centre for Theoretical Physics, University of Oxford, \\
        \small \it 1 Keble Road, Oxford, OX1 3NP, UK \\ \\
        \small $^b$\it Department of Physics and Astronomy, \\
        \small \it University of California, Los Angeles, CA 90095, USA \\ \\
        \normalsize\href{mailto:tomas.andrade@physics.ox.ac.uk}{\texttt{tomas.andrade@physics.ox.ac.uk}}\texttt{, }\href{mailto:sgentle@physics.ucla.edu}{\texttt{sgentle@physics.ucla.edu}}}
\date{}

\begin{document}
\setlength{\baselineskip}{18pt}

\maketitle

\begin{abstract}
\setlength{\baselineskip}{18pt}

Momentum relaxation can be built into many holographic models without sacrificing homogeneity of the bulk solution. In this paper we study two such models: one in which translational invariance is broken in the dual theory by spatially-dependent sources for massless scalar fields and another that features an additional neutral scalar field. We turn on a  charged scalar field in order to explore the condensation of a charged scalar operator in the dual theories. After demonstrating that the relaxed superconductors we construct are thermodynamically relevant, we find that the finite DC electrical conductivity of the normal phase is replaced by a superfluid pole in the broken phase.  Moreover, when the normal phase possesses a Drude behaviour at low frequencies, the optical conductivity of the broken phase at low frequencies can be described by a two-fluid model that is a sum of a Drude peak and a superfluid pole, as was found recently for inhomogeneous holographic superconductors.  We also study cases in which this low-frequency behavior does not hold. We find that the Drude description is accurate when the retarded current-current correlator has 
a purely-dissipative pole  that is well-separated from the rest of the excitations.  

\end{abstract}

\thispagestyle{empty}      

\newpage

\section{Introduction}

Consider a system at finite charge density and temperature.  If momentum is conserved then a non-zero overlap between the electric current and the momentum operator  leads to an infinite DC electrical conductivity.  This feature is indeed observed in the simplest holographic model of such a system: the Reissner-Nordstr\o m-AdS (RN) black brane.   A natural way to build more realistic holographic models that allow momentum to dissipate is to modify the RN geometry such that translational invariance is broken explicitly along the boundary directions.

This can be achieved by introducing spatially-dependent sources for (neutral) matter fields in the bulk.  The backreaction of such sources typically leads to an inhomogeneous solution.  Examples include a spatially-periodic massive neutral scalar \cite{Horowitz:2012ky} or chemical potential \cite{Horowitz:2012gs, Ling:2013nxa, Donos:2014yya}.  In these studies  the DC conductivity is indeed finite, in good agreement with the field theory calculation of \cite{Hartnoll:2012rj}, and the  delta function  is resolved into a Drude peak of the form\footnote{In \cite{Chesler:2013qla}, the spatially-modulated  chemical potential had zero average and a Drude peak in the optical conductivity was not seen.} 
\begin{equation}\label{eq:Drude}
	\sigma(\omega) = \frac{K \tau}{1 - i \omega \tau}
\end{equation}
However, the study of such setups is technically involved because one must solve non-linear PDEs.  It is thus desirable to have a simpler holographic model that incorporates dissipation.\footnote{It was realized early on that the DC conductivity is finite in the probe limit \cite{Karch:2007pd, Faulkner:2010da, Hartnoll:2009ns}; in such a case there exists a large reservoir of neutral matter into which momentum can be transferred. However, considering backreaction reintroduces the delta function.}
In this way one may hope to uncover, in a computationally simple way, universal properties of field theories with holographic duals in which momentum can relax.

One can obtain such a model with a homogeneous bulk geometry by exploiting the presence of a  symmetry in the matter sector.\footnote{One can also exploit a symmetry of the spacetime in order to break boundary translational invariance while retaining homogeneity \cite{Iizuka:2012iv,Donos:2012js}.} For example, a set of massless scalar fields  that are linear in the boundary coordinates were added to the usual Einstein-Maxwell system in \cite{Andrade:2013gsa}. The shift symmetry of the scalars leads to a homogeneous bulk stress tensor. We shall refer to such scalars as axions. Moreover, by choosing the axion configuration in a particular way it is also possible to render the geometry isotropic, allowing for the 
existence of an analytical black brane solution.\footnote{This solution had already been reported in the literature \cite{Bardoux:2012aw}, although 
no holographic applications had been discussed.} 
The finite DC conductivity can be expressed analytically and, in four bulk dimensions,  turns out to be independent of temperature.

At finite chemical potential,  a  transition from a coherent (i.e.\ Drude-like) metallic behaviour to an incoherent one as the parameter that controls the amount of momentum relaxation is increased was identified in this model by \cite{Kim:2014bza}.\footnote{Coherent-incoherent transitions were first observed in holography in \cite{Donos:2012js}. See  \cite{Hartnoll:2014lpa} for a nice introduction to this terminology.} A similar transition was observed in the thermal conductivity of this system at zero chemical potential in \cite{Davison:2014lua}, wherein  this feature was  nicely explained in terms of the structure of the quasi-normal modes (QNM) of the  black brane. There it was observed that the black brane conducts coherently when there is a purely-dissipative mode parametrically separated from the rest of the QNM.

This model can be generalized in a number of ways.  One is to introduce a further neutral scalar field, this time with a potential and couplings to the axions and Maxwell field.  We shall refer to such a scalar as a dilaton.\footnote{For special choices of the coupling functions, such models can  be re-written in terms of neutral complex scalar fields that each have a phase linear in the boundary coordinates.}  Momentum-dissipating black brane solutions to such theories  were constructed  in \cite{Donos:2013eha,Donos:2014uba} and  referred to as `Q-lattices'.  The DC conductivity of these solutions is again finite and determined in terms of the horizon properties of the background \cite{Donos:2014uba, Gouteraux:2014hca}.\footnote{All DC thermoelectric transport coefficients were computed  for these models in \cite{Donos:2014cya}.} Moreover, \cite{Donos:2013eha} identified metallic and insulating phases, the latter of which  being characterized by a vanishing DC conductivity at zero temperature. In the metallic phase,  Drude behaviour is observed for a large range of temperatures, while in the insulating phase the low frequency region is not described by Drude physics. Metal-insulator transitions were observed in similar models in \cite{Donos:2014uba}.\footnote{Metal-insulator transitions were observed recently in an even simpler model by \cite{Mefford:2014gia}.} See \cite{Blake:2014yla} for further studies of such models and also \cite{Taylor:2014tka, Cheng:2014tya} for alternative generalizations.

A conceptually distinct approach was considered in \cite{Vegh:2013sk}, wherein the theory of massive gravity \cite{deRham:2010kj} was employed to remove translational invariance on the boundary as a result of  breaking  (a subset of) diffeomorphisms in the bulk.\footnote{The viability of massive gravity was called into question by \cite{Deser:2014fta}, wherein it was argued that the theory suffers from several pathologies including acausality, superluminality and loss of unique evolution.} The resulting solutions possess a finite DC conductivity for which an analytical expression was provided in \cite{Blake:2013bqa} adapting a calculation in \cite{Iqbal:2008by}. In \cite{Blake:2013owa}, it was argued that there is a qualitative connection between 
massive gravity and a perturbative inhomogeneous lattice generated by a neutral scalar with spatially modulated boundary conditions. 
More precisely, the authors found that, at leading order in the lattice strength,  the equations of motion that govern the conductivity are those of massive gravity if the radially-dependent graviton mass is proportional to the square of the neutral scalar. See \cite{Davison:2013jba, Amoretti:2014zha, Amoretti:2014mma, Baggioli:2014roa, Adams:2014vza} for further studies of massive gravity in this context.

The simplicity of these models makes them amenable to  generalization, allowing us to conveniently study different field theory phenomena in setups in which translational invariance is broken.
The goal of this paper is to study superconductivity in the presence of momentum relaxation, focusing on the axion model and a particular axion-dilaton model. To do so, we shall generalize the by now canonical model of holographic superconductivity of \cite{Hartnoll:2008vx, Hartnoll:2008kx, Gubser:2008px}, consisting of an Einstein-Maxwell-charged scalar theory, by coupling it to the neutral scalar sectors of  \cite{Andrade:2013gsa} and \cite{Donos:2013eha}. The condensation of the charged order parameter is dual to the appearance of charged scalar hair on the black brane (for scale invariant boundary conditions that set the source of the dual operator to zero) as we lower the temperature below a critical value $T_c$. 

Superconducting inhomogeneous lattices were constructed in \cite{Horowitz:2013jaa} in an Einstein-Maxwell-charged 
scalar system with spatially-modulated chemical potential. Below the critical temperature, the optical conductivity
behaves at low frequencies as a two-fluid model consisting of a pole in the imaginary part plus 
a normal Drude component:
\begin{equation}\label{2 fluid model}
	\sigma(\omega) =  \frac{K_n \tau}{1 - i \omega \tau} + i\, \frac{K_s}{\omega} 
\end{equation}
Note that the appearance of a pole in the imaginary part demands there must be a  delta function in the real part with strength set by $K_s$ in order to satisfy the Kramers-Kronig relations.
As the temperature is decreased below $T_c$, it was observed that the parameters $K_s$ and $\tau$ grow rapidly while $K_n$ quickly decreases. 

In addition, other simplified models of momentum relaxation have been utilized in the study of holographic superconductors.  The condensation of a charged scalar was studied in a massive gravity theory in \cite{Zeng:2014uoa}.\footnote{Note that the version of massive gravity considered in \cite{Zeng:2014uoa} only has the mass term 
$\sim \sqrt{g^{\alpha \beta} f_{\alpha \beta}}$ where $f_{\alpha \beta}$ is the reference metric. This is different to the massive gravity theory whose optical conductivity coincides with that of \cite{Andrade:2013gsa}.} Their results qualitatively agree with those in \cite{Horowitz:2013jaa}; namely, they find a two-fluid model with a 
Drude normal component. In  \cite{Koga:2014hwa}, a  charged scalar was coupled to the model of 
\cite{Mateos:2011ix, Mateos:2011tv}, which only has one neutral massless scalar. There it was observed that the critical temperature decreases as the scale of momentum dissipation increases and that a pseudo gap appears in the optical conductivity 
in the direction in which the breaking of translational invariance is introduced. See \cite{Bai:2014poa} for work on the dynamical  
formation of a condensate in the same model. More recently, \cite{Ling:2014laa} studied a holographic superconductor in an axion-dilaton model as we also do below, although with a different 
mass for the dilaton. There it was observed that the critical temperature is reduced by the presence of momentum relaxation and 
that breaking the $U(1)$ from the metallic phase results in a two-fluid model behaviour with Drude normal component. This is not so in the insulating phase, for which the normal component is not well described by a Drude peak. See also \cite{Aprile:2014aja} for an alternative approach.

In this paper we construct superconducting solutions and observe the transition from all the unbroken phases we have considered; namely, 
the exact solution of \cite{Andrade:2013gsa, Bardoux:2012aw} in the axion model and one choice of metallic and insulating solutions 
in the axion-dilaton model of \cite{Donos:2013eha}. Crucially, we first demonstrate that the broken phases are indeed  
thermodynamically preferred, confirming their physical relevance. 

Our main result is the observation of a zero-frequency pole in the imaginary part of the electrical conductivity, confirming that our solutions are indeed superconducting. Moreover, in cases in which the normal phase possesses a Drude behaviour at low frequencies (namely, in the axion model with small momentum dissipation parameter and the metallic solution in the axion-dilaton model), the broken phase can be described as a two-fluid model whose low-frequency conductivity is given by  \eqref{2 fluid model}. 
We find that this can be explained by the structure of the lowest QNMs: Drude peaks arise when there is a purely-dissipative mode that is well-separated from the rest of the excitations, resembling
the picture of  \cite{Davison:2014lua}. When this ceases to be true, we observe departures from the Drude behaviour in the real part of the conductivity and thus the first term in \eqref{2 fluid model} must be modified.
We study the temperature dependence of the parameters in \eqref{2 fluid model} for the broken phase of the metal of the axion-dilaton model. As we lower the temperature from $T_c$,  we find that 
$\tau$ first decreases slowly and then increases quickly, $K_n$ is monotonically decreasing reaching a small value, 
and $K_s$ monotonically increases approaching a constant. We find that $K_s$ exhibits the same behaviour in the broken phase of the insulator and also in the broken phase studied in the axion model. 

We also check the Ferrell-Glover-Tinkham (FGT) sum rule \cite{Ferrell:1958zza, PhysRevLett.2.331}, which relates `missing' spectral weight to the strength of the zero-frequency pole. We verify that the sum rule holds for the broken metallic phase and the broken phase in the axion 
model, as was also demonstrated for the inhomogeneous lattice of \cite{Horowitz:2013jaa}.

This paper is organized as follows.  In section~\ref{sec:axion} we study the axion model coupled to a charged scalar.  
We describe a general class of solutions which implement momentum relaxation and discuss the calculation of the optical conductivity 
for such configurations.  
After a review of the normal phase and its conductivity, we move on to explore the broken phase. In section~\ref{sec:axion_dilaton} 
we also follow this structure for a particular axion-dilaton model and study both metallic and insulating states.   We conclude 
with a discussion of our results and comments on extensions and more general models in section~\ref{sec:Discussion}.

\section{Superconductivity in an axion model}\label{sec:axion}

Both models we will study have the form
\begin{equation}
I = I_{\textrm{N}} +I_{\textrm{C}} \equiv \int d^4x\sqrt{-g}\, \mathcal{L}
\end{equation}
In this section we focus on an axion model that has neutral sector \cite{Andrade:2013gsa}
\begin{equation}
	I_{\textrm{N}} = \int d^4 x \sqrt{-g} \left[ R + 6 - \frac{1}{4} F^2 - \frac{1}{2}  \sum_{i=1}^2 (\partial \chi_i)^2  \right ] \quad \textrm{where}\quad F=dA
\end{equation}
The field equations admit AdS$_4$ with unit radius as a vacuum solution.  We add to this action a charged sector given by
\begin{equation}\label{eq:chargedaction}
	I_{\textrm{C}} = \int d^4x \sqrt{-g} \left[ -  | D \psi  |^2 - m_{\psi}^2 |\psi|^2  \right ] 
\quad \textrm{where}\quad
	D_a \psi = \partial_a \psi - i q A_a \psi
\end{equation}
We will always choose this simple potential with $m_{\psi}^2=-2$.  The charged scalar $\psi$ is dual to a scalar operator $\mathcal{O}$ of dimension $\Delta=2$ charged under a global $U(1)$.  This model contains the minimal ingredients needed to construct a holographic superconductor in the presence of dissipation. Our main goal here is to study the optical conductivity in the broken phase of this model.

Both the normal phase and broken phase are described by black brane solutions. We make the following isotropic ansatz 
for the bulk fields:
\begin{gather}
	ds^2 = - U(r) dt^2  + \frac{dr^2}{U(r)} + e^{2V(r)} (dx^2 + dy^2) \\
	A = A_t(r) dt,  \quad \chi_1 = \alpha x, \quad \chi_2 = \alpha y,  \quad \psi = \psi(r)
\end{gather}
Note that the metric is isotropic and homogeneous but the full solution is not due to the presence of the spatial dependence 
in the axion fields $\chi_i$. The parameter $\alpha$ is the strength of the translational symmetry breaking in the dual theory, and setting
$\alpha = 0$ restores translational invariance.

In order to work at non-zero temperature in the field theory we demand a regular non-degenerate horizon at $r=r_+$.   This implies $U(r_+)=A_t(r_+)=0$ and that the temperature is given by
\begin{equation}
T = \frac{U'(r_+)}{4 \pi} = \frac{1}{16 \pi V'(r_+)} \left(12-A_t'(r_+)^2-2 \alpha^2  e^{-2 V(r_+)}+4 \psi(r_+)^2\right)
\end{equation}
We find the following asymptotic expansion near the AdS$_4$ boundary at $r \to \infty$: 
\begin{align}
A_t &= \mu - \frac{\rho}{r}+ \ldots \label{eq:axion_gauge_field}\\
\psi &= \frac{\psi_1}{r} + \frac{\psi_2}{r^2} + \ldots \\
U &= r^2 -\frac{\alpha^2}{2} -\frac{m}{r}+\ldots\\
V & = \log r+ \ldots 
\end{align}
We consider the standard quantization of the charged scalar, in which $\psi_1$ is  interpreted as the source and $\psi_2$ as the expectation value of $\mathcal{O}$. Thus, a bulk solution with $\psi_2 \neq 0$ and $\psi_1 = 0$ corresponds to the broken phase in the dual theory. Such solutions  have been thoroughly investigated in the literature in the special case of $\alpha = 0$, but in section~\ref{sec:brokenaxion} we turn on $\alpha \neq 0$,  allowing us to study  the spontaneous breaking of the $U(1)$ symmetry in the presence of dissipation.  The remaining parameters in this expansion have the following interpretation in the dual theory: $\mu$ is the chemical potential, $\rho$ is the total charge density and $m$ is proportional to the total energy density.  The phase space of the dual theory is parametrized by the dimensionless quantities $T/\mu$ and $\alpha/\mu$.

In order to compute the optical conductivity, we consider fluctuations of the gauge field of the form $\delta A_x = e^{-i\omega t}a_x(r)$. This couples to fluctuations of the metric and the axion fields, which we write as
\begin{equation}
	\delta g_{tx} = e^{-i\omega t} h_{tx}(r), \qquad  \delta \chi_1 = e^{-i\omega t} \ss (r)
\end{equation}
The equations that govern these fluctuations are
\begin{align}
	\ss'' + \left( 2 V' + \frac{U'}{U} \right) \ss'+  \frac{\omega^2}{U^2} \ss - i \omega \alpha  \frac{e^{- 2 V} h_{tx}}{U^2}  &= 0 \\
    a_x'' + \frac{U'}{U} a_x' + \left(  \frac{\omega^2}{U^2} - \frac{2 q^2 \psi^2}{U}  \right) a_x + \frac{A_t'}{U} h_{tx}' 
    - \frac{2 A_t' V'}{U} h_{tx} &=0 \\
	\alpha \ss - \frac{i \omega}{U} (  A_t' a_x + h_{tx}' - 2 V' h_{tx}  ) &= 0
\end{align}
Note that the only difference when a charged scalar $\psi$ is turned on is an additional mass term for $a_x$. 
Near the conformal boundary, the linearized fields behave as
\begin{align}
	\ss &= \ss^{(0)} + \frac{\omega}{2 r^2} (\omega \ss^{(0)} - i \alpha h_{tx}^{(0)})  + \frac{\ss^{(3)}}{r^3} + \ldots \\
	a_x &= a_x^{(0)} + \frac{a_x^{(1)}}{r}  + \ldots \\
	h_{tx} &= r^2 h_{tx}^{(0)} + \ldots 
\end{align}
The optical conductivity is defined via
\begin{equation}\label{eq:sigmadef}
\sigma(\omega) = \frac{G^R_{J^x J^x}}{i\omega}= \frac{a_x^{(1)}}{i \omega a_x^{(0)}}
\end{equation}
where $G^R_{J^x J^x}$ is the retarded Green's function for the current operator $J^x$ in the dual theory.
In order to compute it, we need to turn on the source $a_x^{(0)}$ with all other sources set to zero.  
This can be achieved in a gauge-invariant way by requiring (see e.g.\ \cite{Donos:2013eha})
\begin{equation}
	\omega \ss^{(0)} - i \alpha h_{tx}^{(0)} = 0
\end{equation}
In addition, to obtain the retarded correlator the linearized fields must satisfy ingoing boundary conditions at the horizon, 
which amounts to 
\begin{align}\label{ingoing bc axion 1}
	\ss &= (r - r_+)^{- i \omega/(4 \pi T)} ( \ss^{(+)} + \ldots  ) \\
\label{ingoing bc axion 2}
	a_x &= (r - r_+)^{- i \omega/(4 \pi T)} ( a_x^{(+)} + \ldots  ) \\
\label{ingoing bc axion 3}
	h_{tx} &= (r - r_+)^{- i \omega/(4 \pi T)} ( h_{tx}^{(+)} (r-r_+) + \ldots  )
\end{align}
\noindent where the ellipses denote regular power series in $(r-r_+)$. The leading order coefficients satisfy the relation
\begin{equation}
	\left( \frac{i \omega}{4 \pi T} - 1 \right) h_{tx}^{(+)} =   A_t'(r_+) a_x^{(+)} + \alpha \ss^{(+)} 
\end{equation}
Imposing these conditions, we use a shooting method to read off the optical conductivity.

As mentioned in the introduction, it will often be useful to study the quasi-normal frequencies of the black brane associated with such fluctuations.  These are given by the  poles in $G^R_{J^x J^x}$ \cite{Birmingham:2001pj, Son:2002sd}, which from  \eqref{eq:sigmadef}  are located at frequencies for which $a_x^{(0)}=0$.  

Before moving forward to discuss our results for the broken phase, first we review known results for both the DC and AC conductivity in the normal phase.

\subsection{Normal phase}

We choose the following  black brane solution  as  the normal phase \cite{Andrade:2013gsa, Bardoux:2012aw}:
\begin{equation}\label{eq:baldaxionbrane}
\begin{gathered}
U(r)= r^2  - \frac{\alpha^2}{2}- \frac{m}{r} + \frac{\mu^2 }{4} \frac{ r_+^2 }{ r^2 }, \quad V(r) = \log r \\
 A_t(r) = \mu \left(1 - \frac{r_+}{r} \right), \quad \psi(r)=0
\end{gathered}
\end{equation}
The parameter $m$ is determined by  solving $U(r_+)=0$.   The temperature of the black brane is simply
\begin{equation}
T = \frac{U'(r_+)}{4 \pi} = \frac{1}{16 \pi r_+} \left( 12 r_+^2   -  \mu^2  - 2 \alpha^2 \right)
\end{equation}
It is clear that we recover the RN solution by setting $\alpha=0$.

The DC conductivity in this phase was shown in \cite{Andrade:2013gsa} to take the temperature-independent value
\begin{equation}\label{eq:axion_sigmaDC}
\sigma_{\textrm{DC}} = 1 + \frac{\mu^2}{\alpha^2}
\end{equation}
The AC conductivity was studied in detail in \cite{Kim:2014bza}. Its real part  interpolates between the DC value at zero frequency and unity at high frequency with a minimum at intermediate frequencies whose depth and location depends on $T/\mu$ and $\alpha/\mu$.  The imaginary part has a peak at low frequencies then goes to zero as the frequency goes to zero.  
It was argued in  \cite{Kim:2014bza} that this behaviour is well-described by a slight modification of the Drude 
form \eqref{eq:Drude} when $\alpha/\mu$ is small but not when $\alpha/\mu$ is large.  Indeed, we checked that when $\alpha$ is much larger than both $T$ and $\mu$, the conductivity is close to unity for all frequencies and there is clearly no coherent Drude peak.

\subsection{Broken phase}\label{sec:brokenaxion}

Next we study solutions with the charged scalar turned on.  We construct these numerically using a shooting method.  At all $\alpha/\mu$ studied we find a black brane solution with charged scalar hair that exists below some critical $T_c/\mu$.   In figure~\ref{fig:axion_Tc} we show this critical temperature as a function of $\alpha/\mu$ for different values of the scalar charge $q$. Note that the curves become non-monotonic as $q$ is increased, in contrast to several studies mentioned in the introduction.  From now on we fix $q=2$.  In figure~\ref{fig:axion_condensate} we show a family of condensate curves for different $\alpha/\mu$.
\begin{figure}[!ht]
\begin{center}
\includegraphics[width=0.5\textwidth]{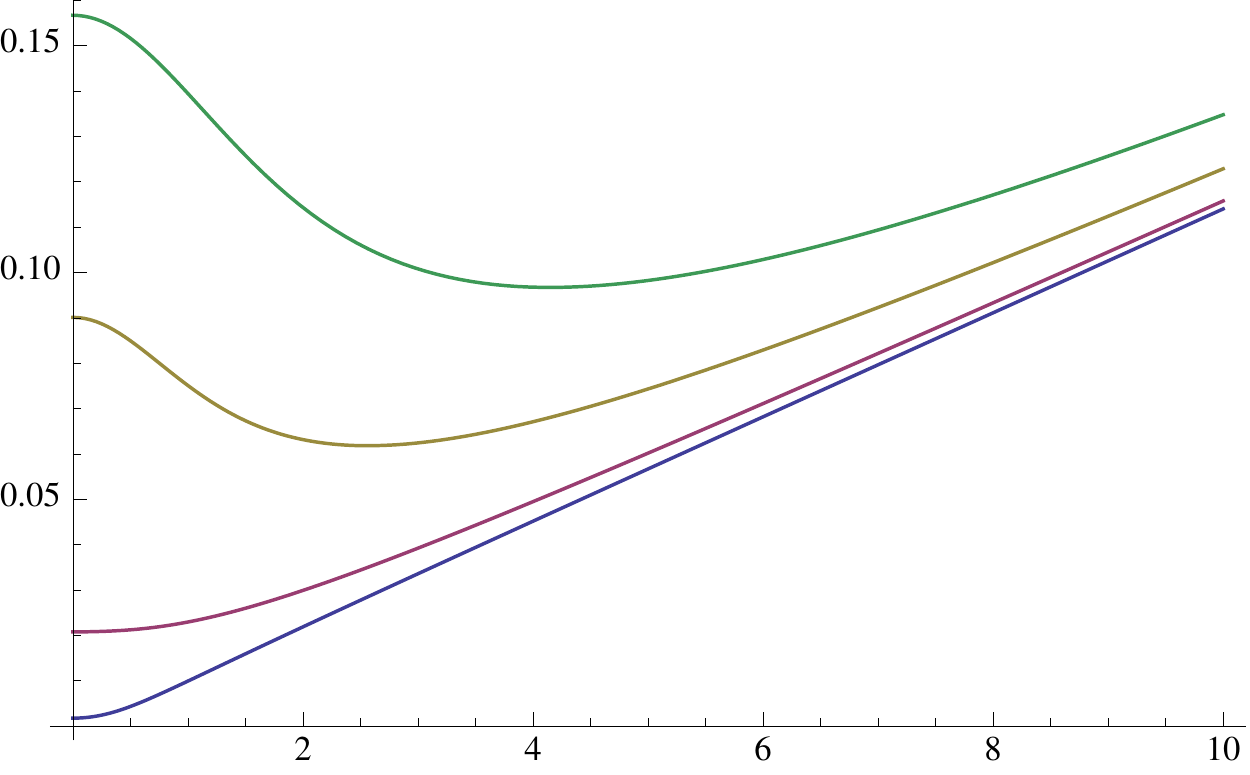}
\setlength{\unitlength}{0.1\textwidth}
\begin{picture}(0.3,0.4)(0,0)
\put(-5.7,2.5){\makebox(0,0){$\frac{T_c}{\mu}$}}
\put(0.2,0.2){\makebox(0,0){$\frac{\alpha}{\mu}$}}
\end{picture}
\end{center}
\vskip-1em
\caption{Critical temperature as a function of $\alpha/\mu$ for $q=0.5,1,2,3$ (labelled from bottom to top).}\label{fig:axion_Tc} 
\vskip-1em
\end{figure}
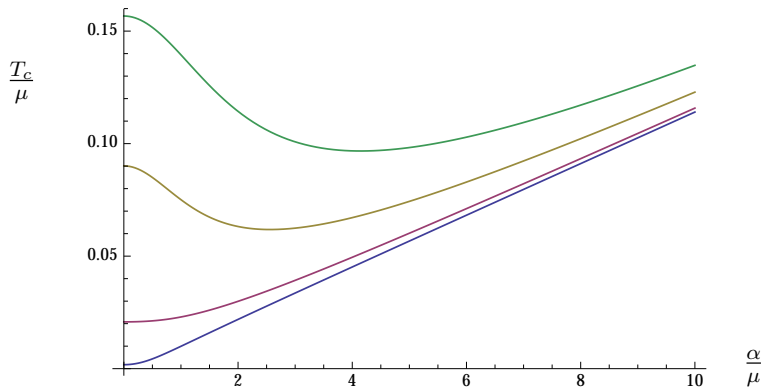
\begin{figure}[!ht]
\begin{center}
\includegraphics[width=0.5\textwidth]{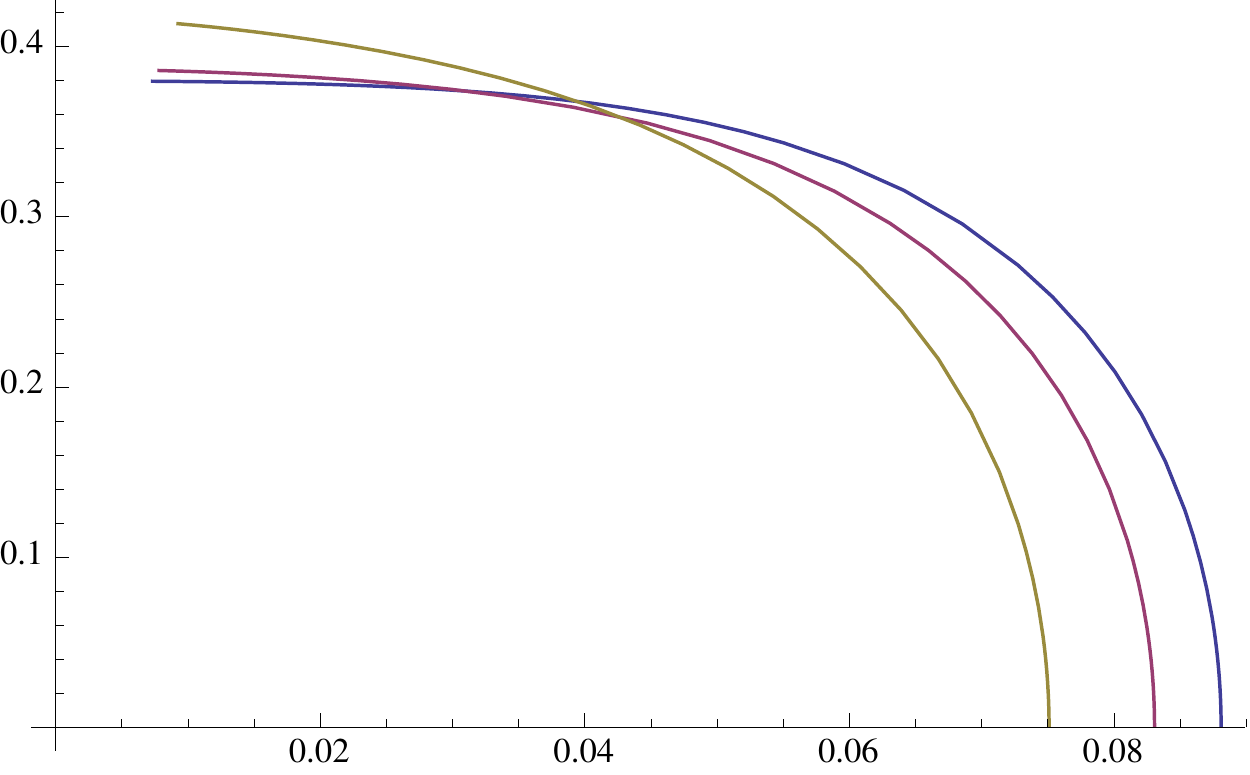}
\setlength{\unitlength}{0.1\textwidth}
\begin{picture}(0.3,0.4)(0,0)
\put(-5.7,2.5){\makebox(0,0){$\frac{\psi_2}{\mu^2}$}}
\put(0.2,0.2){\makebox(0,0){$\frac{T}{\mu}$}}
\end{picture}
\end{center}
\vskip-1em
\caption{Condensate as a function of temperature for $\alpha/\mu=0.3, 0.6, 1$ (labelled from right to left).}\label{fig:axion_condensate} 
\end{figure}
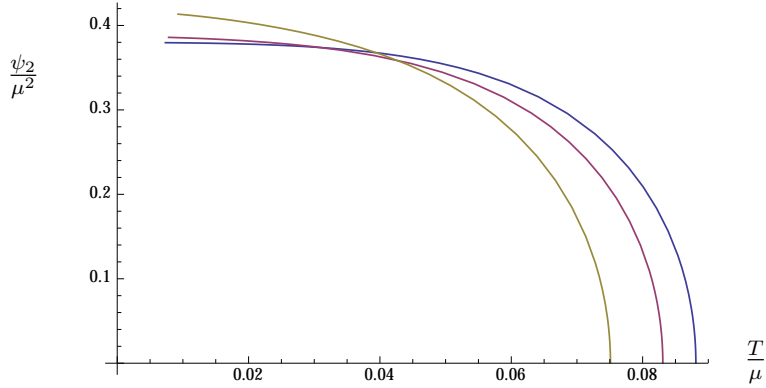

\begin{figure}[!ht]
\begin{center}
\includegraphics[width=0.5\textwidth]{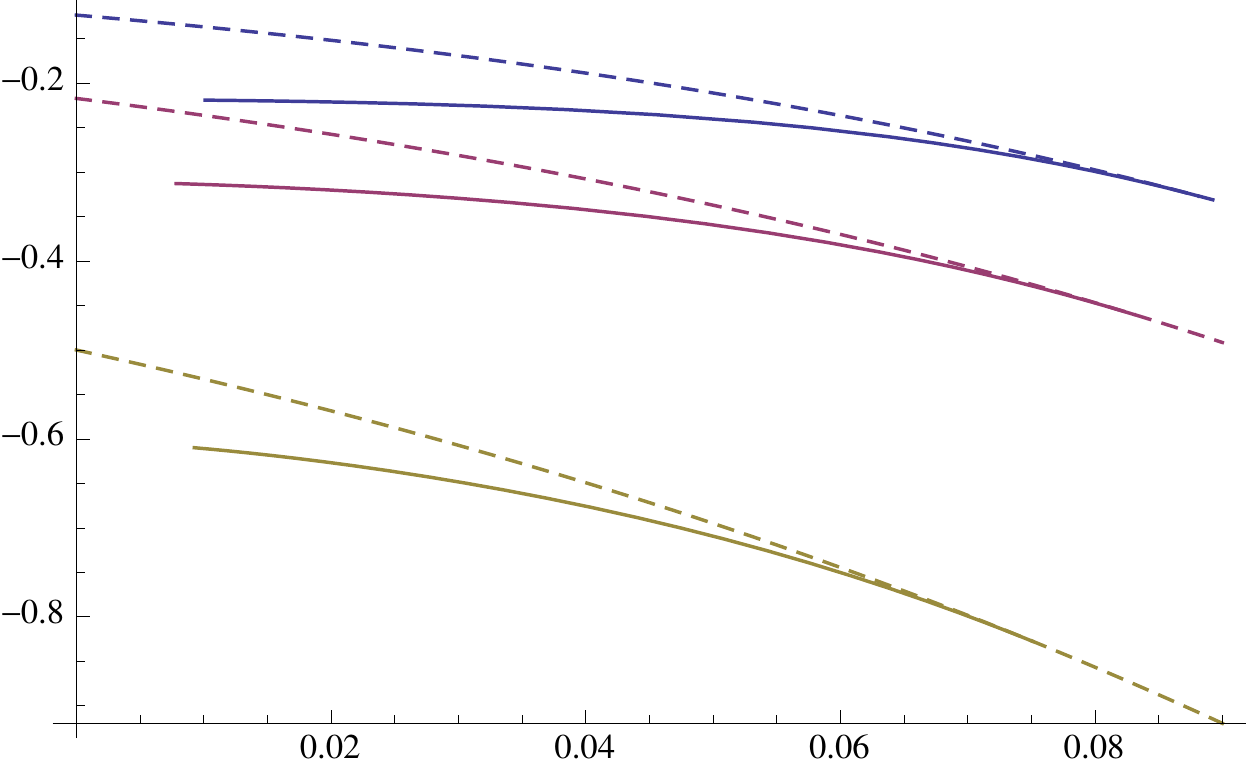}
\setlength{\unitlength}{0.1\textwidth}
\begin{picture}(0.3,0.4)(0,0)
\put(-5.7,2.5){\makebox(0,0){$\frac{w}{\mu^3}$}}
\put(0.2,0.2){\makebox(0,0){$\frac{T}{\mu}$}}
\end{picture}
\end{center}
\vskip-1em
\caption{Free energy density as a function of temperature for the broken phase (solid lines) and the normal phase (dashed lines) at  $\alpha/\mu=0.3, 0.6, 1$ (labelled from top to bottom).}\label{fig:axion_free_energy} 
\end{figure}
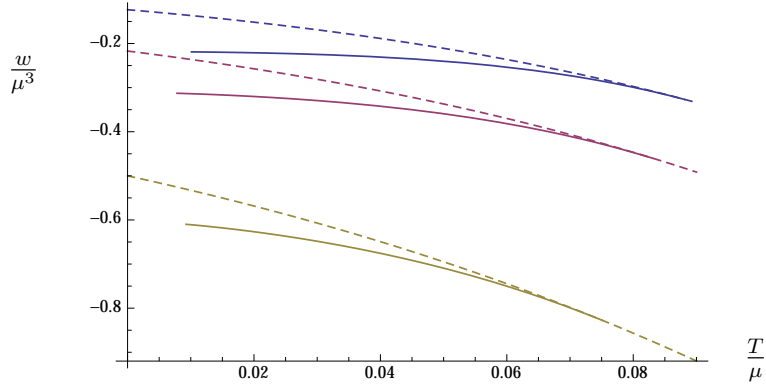

In order to determine whether these solutions are thermodynamically relevant, we must compute their free energies.  First we Wick rotate to Euclidean signature via 
\begin{equation}
t = - i \tau \quad \textrm{and}\quad I_E = - i I
\end{equation}
The free energy density $w$ is given by the renormalised on-shell Euclidean action via
\begin{equation}
I_E^{\textrm{ren}} = \beta V_2 w
\end{equation}
where $\tau \sim \tau + \beta$ and $\beta=1/T$. Using Einstein's equations we can write the on-shell Lagrangian as a total derivative:
\begin{align}
I_E &= -\int d^4x \sqrt{g}\, \mathcal{L} = \int d^4x \left[ 2 \sqrt{g}\, U\, V' +\alpha^2 r \right]' \nonumber \\
      & = \beta V_2 \left[ \lim_{r\to\infty} (2\sqrt{g}\, U\, V' +\alpha^2 r) - \alpha^2 r_+\right]
\end{align}
where in the last line we have used $U(r_+) = 0$.  We must supplement the action with boundary terms in order to have a well-defined variational principle and also to obtain a finite $w$ \cite{Henningson:1998gx, Balasubramanian:1999re}. We find that\footnote{Our conventions are as follows:  $h_{ab} = g_{ab}+n_an_b$ is the induced metric on the timeline boundary at large $r$ with outward-pointing unit normal $n_a$ and  $\mathcal{K}_{ab} = h_{(a}^c h_{b)}^d\nabla_{c} n_{d}$ is the extrinsic curvature with trace $\mathcal{K} = h^{ab}\mathcal{K}_{ab} $.}
\begin{equation}
I_E^{\textrm{ren}} = I_E + \int_{r\to\infty} d^3x \sqrt{h}\left[ -2 \mathcal{K}+4 - \frac{1}{2}\sum_{i=1}^2 (\partial\chi_i)^2+|\psi|^2\right] 
\end{equation}
yields the result
\begin{equation}
w = -(m + r_+ \alpha^2)
\end{equation}
Note that this formula is valid for both the  normal and broken phases, but $m$ is found numerically for the latter.   In figure~\ref{fig:axion_free_energy} we demonstrate that the broken phase has lower free energy density than the normal phase at the same $\alpha/\mu$ and thus is thermodynamically preferred.

Now we present our results for the optical conductivity in this phase.  In figure~\ref{fig:axion_sigma} we show our results for $\alpha/\mu =0.3$ for a range of temperatures below $T_c/\mu$ and in figure~\ref{fig:axion_sigma_log} we plot the same data on a log-log scale.  Two main features are apparent.  Firstly, the imaginary part of the conductivity diverges at the origin. By fitting the imaginary part at small frequencies we confirm the existence of a $1/\omega$ pole, which leads to a delta function at the origin in the real part via the Kramers-Kronig relations. 
\begin{figure}[!ht]
\begin{center}
\hskip1em
\includegraphics[width=0.4\textwidth]{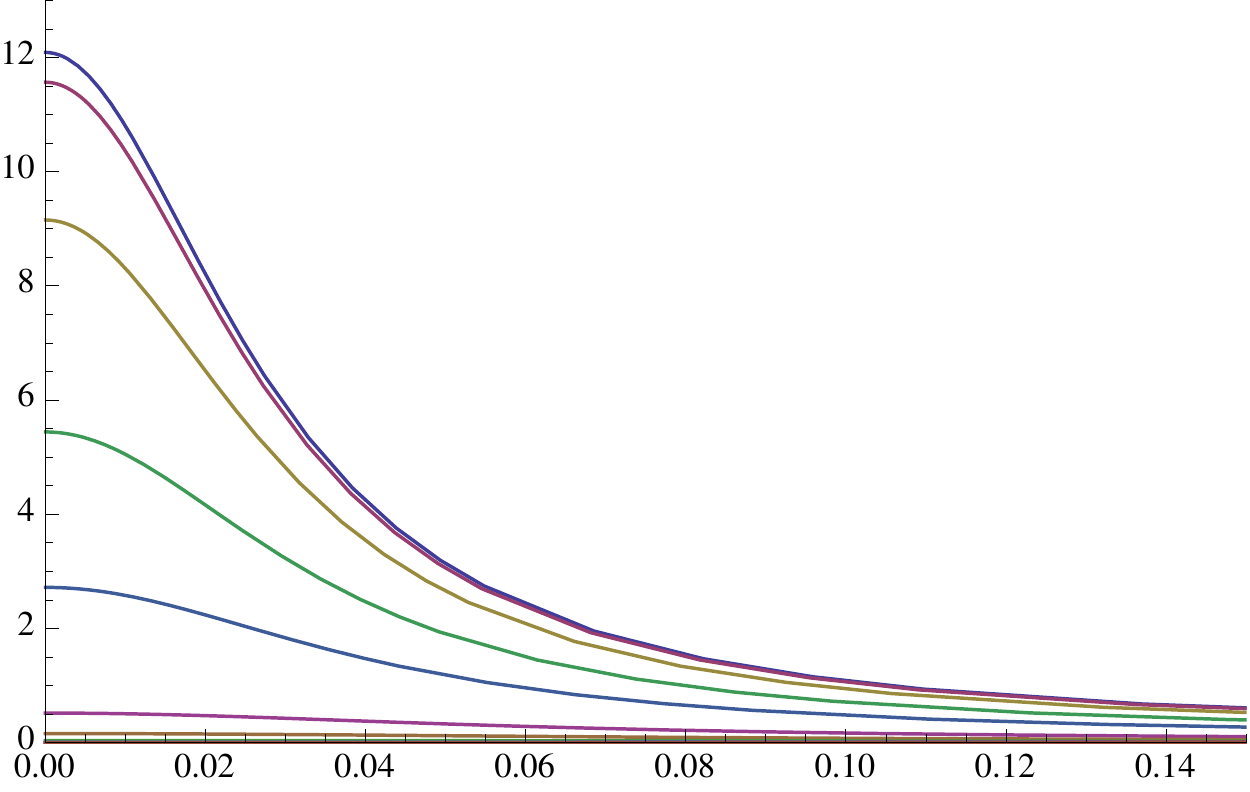}
\hskip2em
\includegraphics[width=0.4\textwidth]{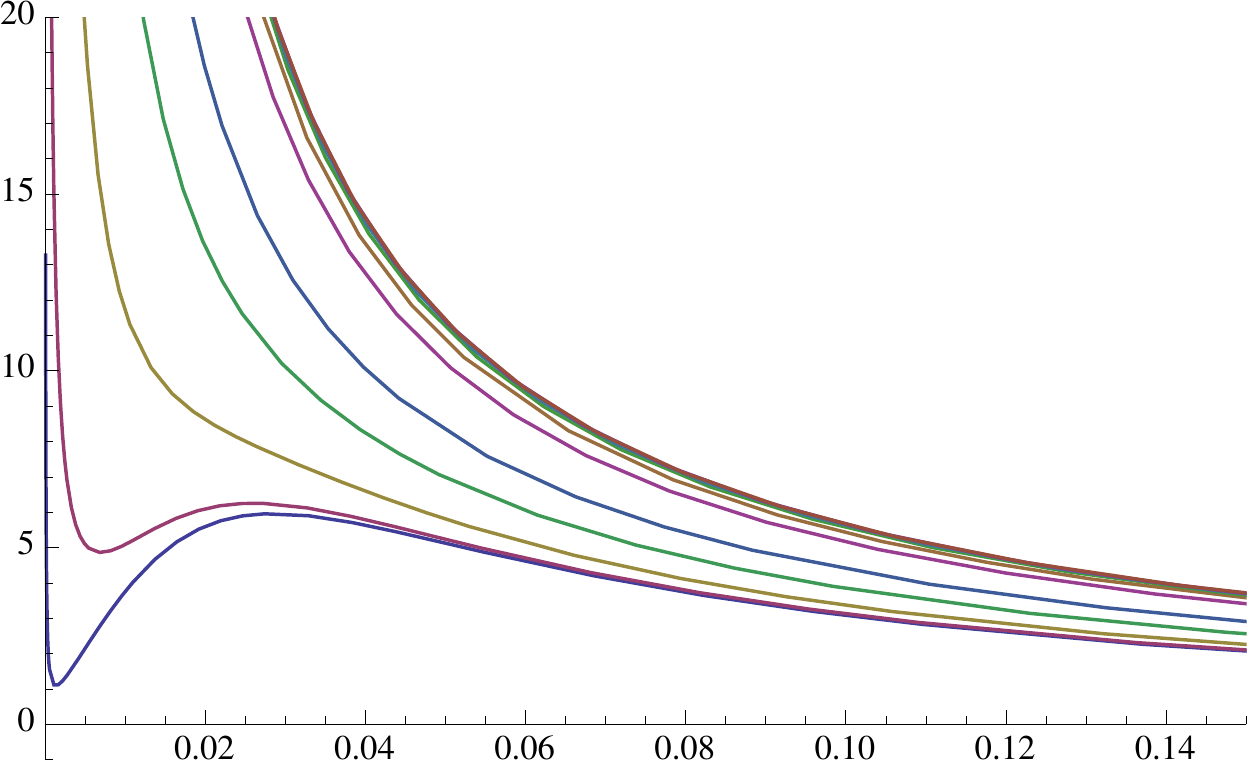}
\setlength{\unitlength}{0.1\textwidth}
\begin{picture}(0.3,0.4)(0,0)
\put(-9.1,2.){\makebox(0,0){$\Re\sigma$}}
\put(-4.6,2.){\makebox(0,0){$\Im\sigma$}}
\put(0.2,0.2){\makebox(0,0){$\frac{\omega}{\mu}$}}
\end{picture}
\end{center}
\vskip-1em
\caption{Conductivity in  the broken phase at $\alpha/\mu = 0.3$ as a function of frequency  for various temperatures. 
Temperature decreases from $T/\mu\lesssim T_c/\mu=0.088$ down to $0.0073$, from top to bottom in the left plot and from bottom to top in the right.}\label{fig:axion_sigma} 
\end{figure}
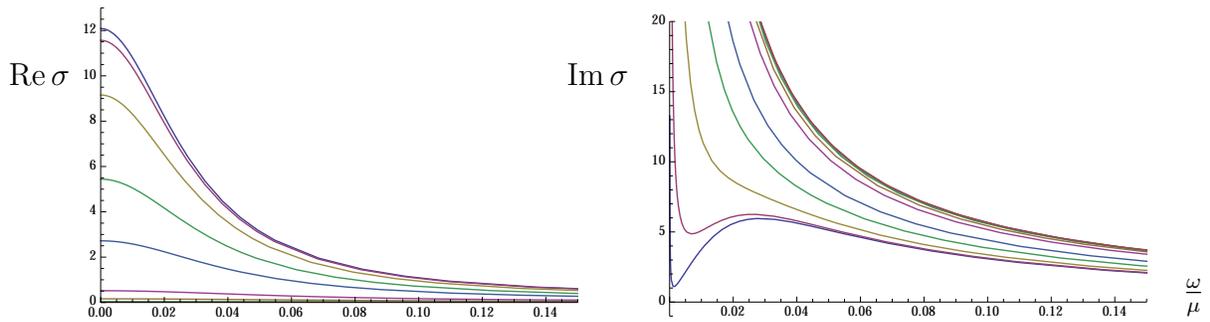
\begin{figure}[!ht]
\vskip-0.5em
\begin{center}
\hskip1em
\includegraphics[width=0.4\textwidth]{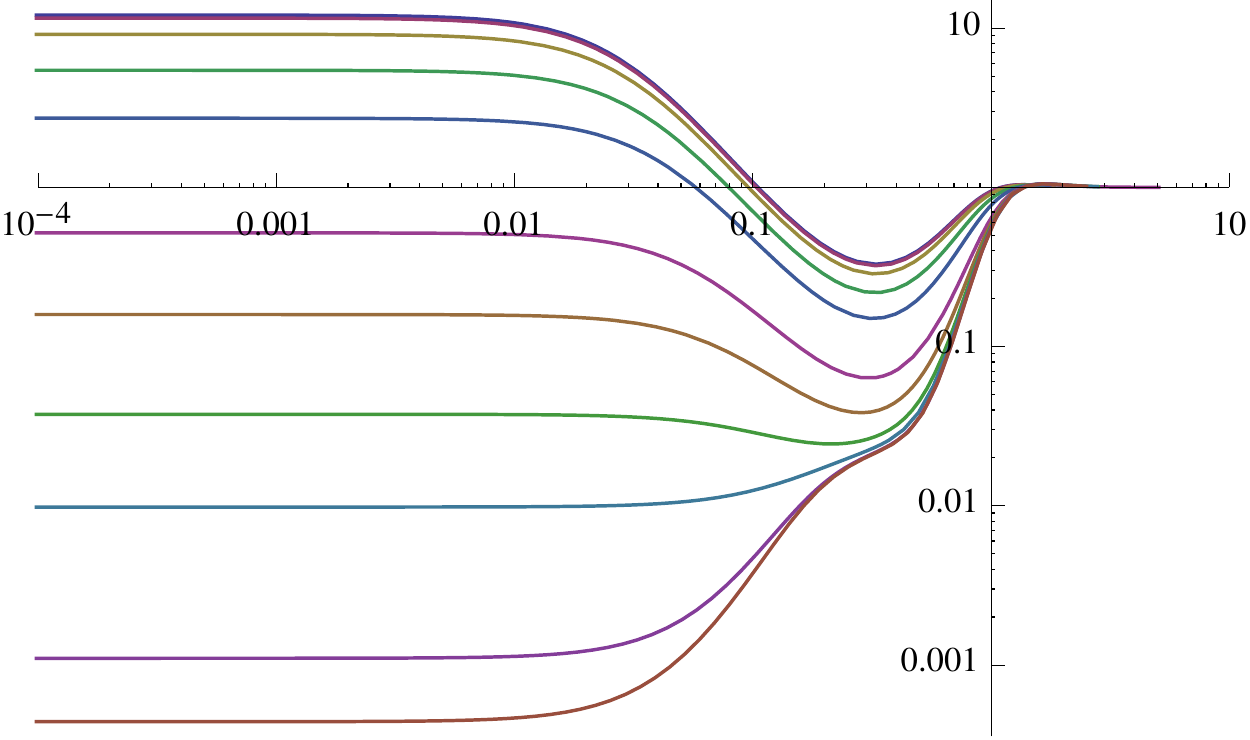}
\hskip2em
\includegraphics[width=0.4\textwidth]{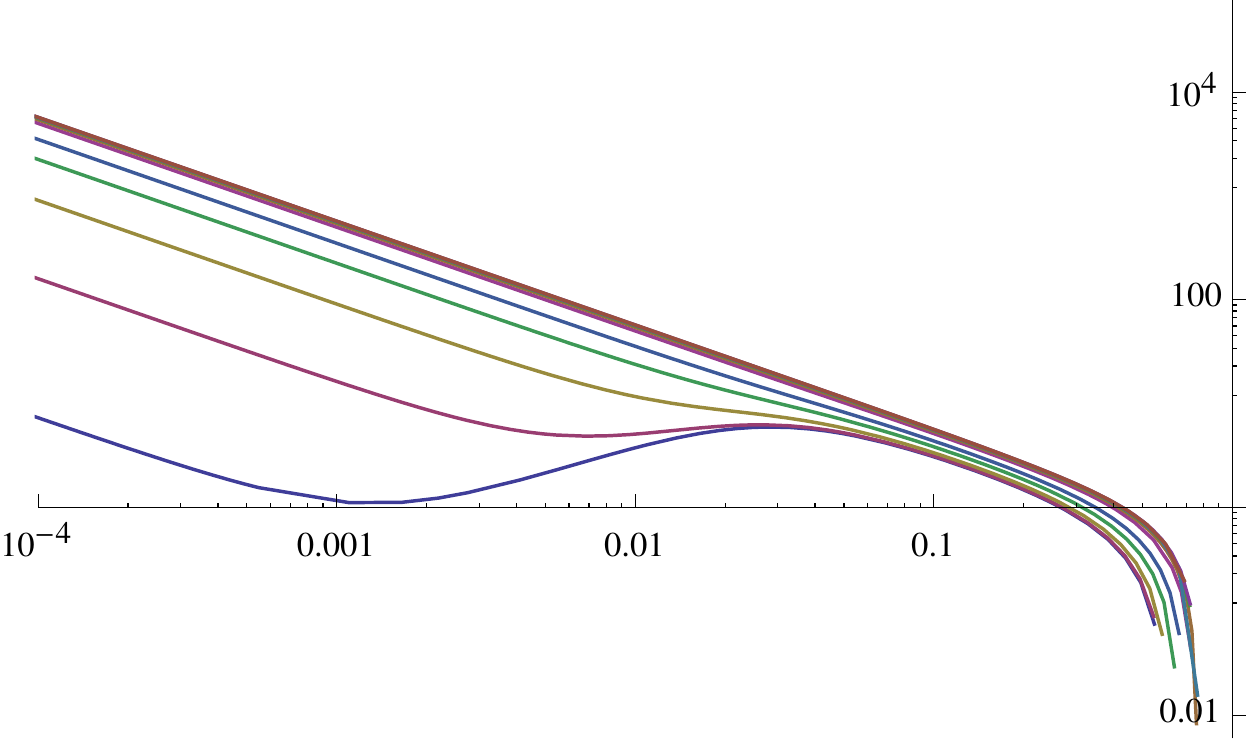}
\setlength{\unitlength}{0.1\textwidth}
\begin{picture}(0.3,0.4)(0,0)
\put(-9.1,2.){\makebox(0,0){$\Re\sigma$}}
\put(-0.8,2.){\makebox(0,0){$\Im\sigma$}}
\put(0.2,0.7){\makebox(0,0){$\frac{\omega}{\mu}$}}
\end{picture}
\end{center}
\vskip-1.5em
\caption{The same data as in figure~\ref{fig:axion_sigma} plotted on a log-log scale.}\label{fig:axion_sigma_log} 
\vskip-0.5em
\end{figure}
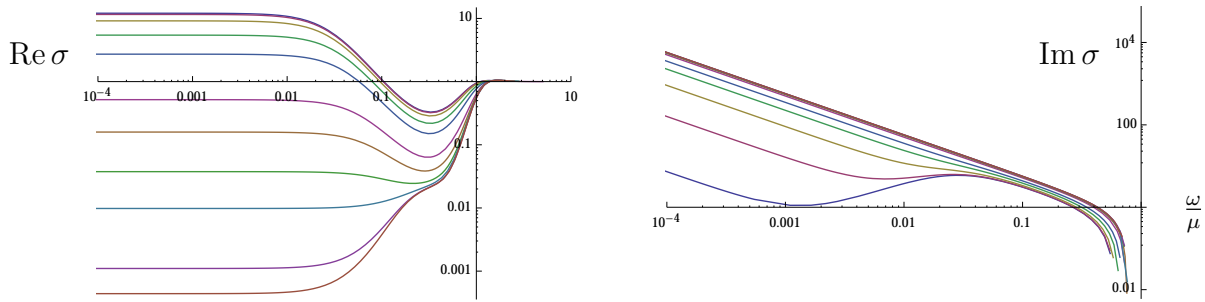

In fact,  at temperatures close to $T_c/\mu$ we find that the low frequency region of the conductivity can be well-approximated by 
a two-fluid model with Drude normal component given by \eqref{2 fluid model}.  Specifically, we fit the real part of our data to the real part of this formula, then do the same for the imaginary counterparts over the same frequency range.  For a temperature $T/\mu\lesssim T_c/\mu=0.088$ with $\omega/\mu<0.05$ we find that the squared sum of the residuals for these fits is $7.8 \times 10^{-3}$ or $2.2\times 10^{-4}$, respectively. We also find that the value of  $1/(\tau \mu)$ extracted from the real fit differs by that from the imaginary fit by $1.9 \%$ and from the frequency of the lowest purely-dissipative QNM by $1.4 \%$. 

\begin{figure}[!ht]
\begin{center}
\includegraphics[width=0.5\textwidth]{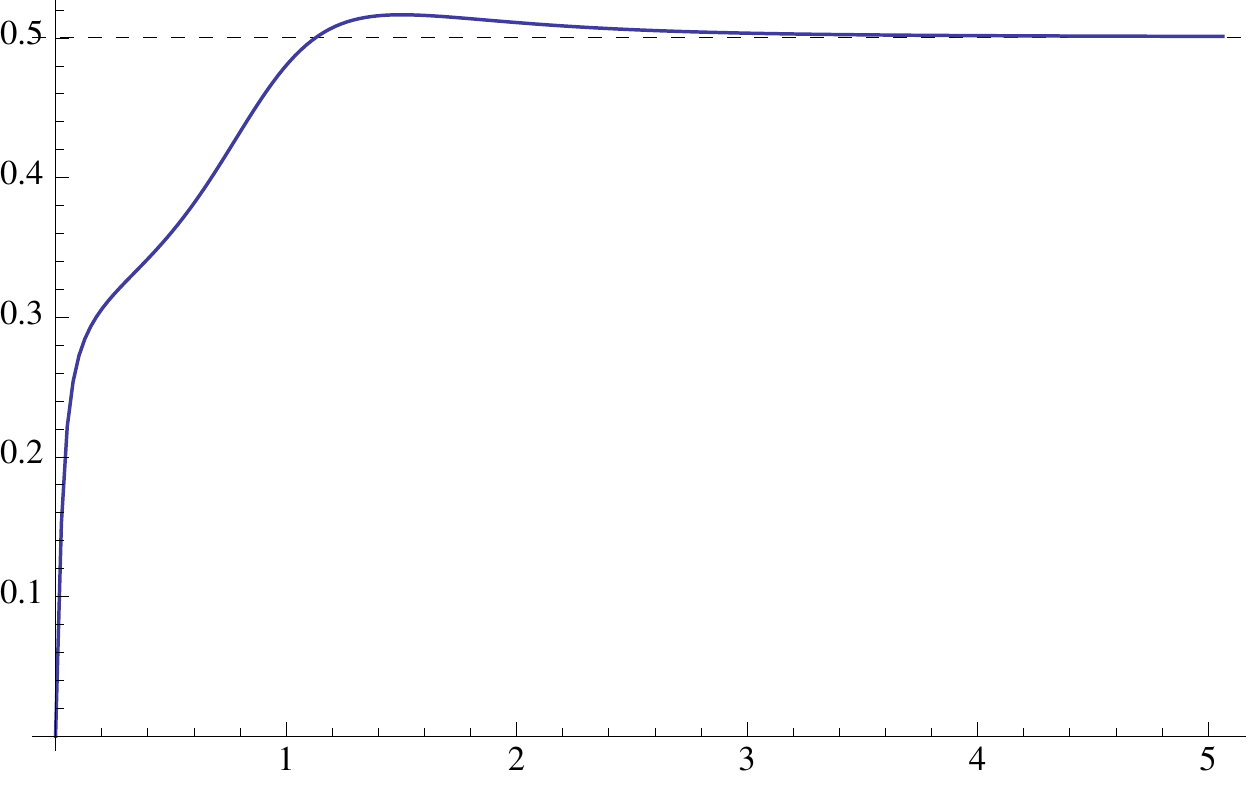}
\setlength{\unitlength}{0.1\textwidth}
\begin{picture}(0.3,0.4)(0,0)
\put(-5.7,2.5){\makebox(0,0){$F$}}
\put(0.2,0.2){\makebox(0,0){$\frac{\omega}{\mu}$}}
\end{picture}
\end{center}
\vskip-1em
\caption{Confirmation of the FGT sum rule \eqref{eq:FGT} at  $\alpha/\mu=0.3$ where we choose a broken phase  at $T/\mu=0.62\, T_c/\mu$.  The dashed line is the value of $K_s/\mu=0.50$ extracted from a fit of $\Im\sigma_s$.}\label{fig:axion_sum_rule} 
\end{figure}
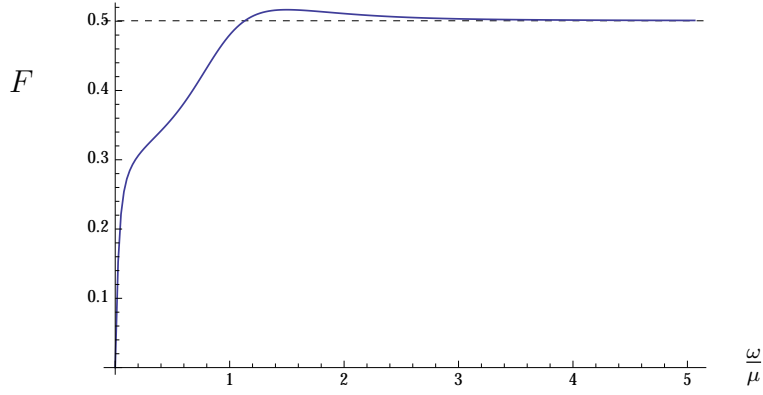

Secondly, in the real part of the conductivity at low frequencies we see a departure from the normal phase value \eqref{eq:axion_sigmaDC}  as the temperature is lowered.  This is a result of  spectral weight being transferred into the superfluid component of the fluid.  The coefficient of the zero-frequency pole measures the superfluid density and is governed by the Ferrell-Glover-Tinkham sum rule:
\begin{equation}\label{eq:FGT}
\frac{K_s}{\mu} = \frac{2}{\pi}\lim_{\omega/\mu\to\infty} F(\omega/\mu)\quad\textrm{with}\quad F(p)\equiv \int_{0^+}^{p} dp' \, \Re \left[\sigma_n(p') - \sigma_s(p')\right]
\end{equation}
where $\sigma_n$ is the  conductivity of the normal phase at $T_c/\mu$ and $\sigma_s$ is the conductivity of the broken phase at some $T/\mu<T_c/\mu$.  In figure~\ref{fig:axion_sum_rule} we demonstrate that, going to high enough frequencies, $F(\omega/\mu)$ does indeed tend to the coefficient of the $\frac{1}{\omega/\mu}$ pole extracted from a fit of $\Im\sigma_s(\omega/\mu)$ at low frequency.  Such agreement was also found for the inhomogeneous model studied in \cite{Horowitz:2013jaa}, where it is connected to the missing spectral weight observed in some cuprates.  We discuss the interpretation of $K_s$ in section~\ref{sec:Discussion}.

As the temperature is lowered, the peak at low frequencies in the imaginary part of the conductivity is overwhelmed by the superfluid pole.  This is not in conflict with \eqref{2 fluid model}; rather, it is simply a result of the interplay between the $1/\omega$ and $\omega$ components.  However, one can see from figure~\ref{fig:axion_sigma_log} that for very low temperatures the local maximum in the real part at low frequencies also disappears.  One might have expected to see the maximum reappear at lower frequencies but we think it is unlikely we have missed such a feature since our temperature steps are quite small.  

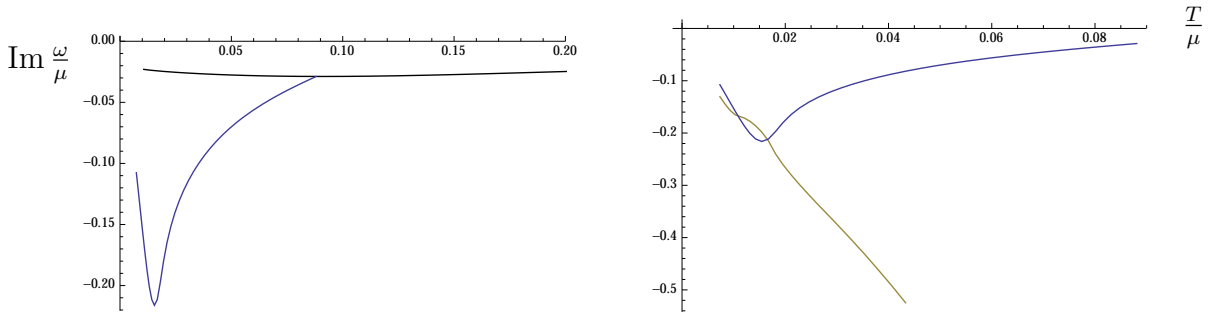
\begin{figure}[!h]
\begin{center}
\hskip1em
\includegraphics[width=0.4\textwidth]{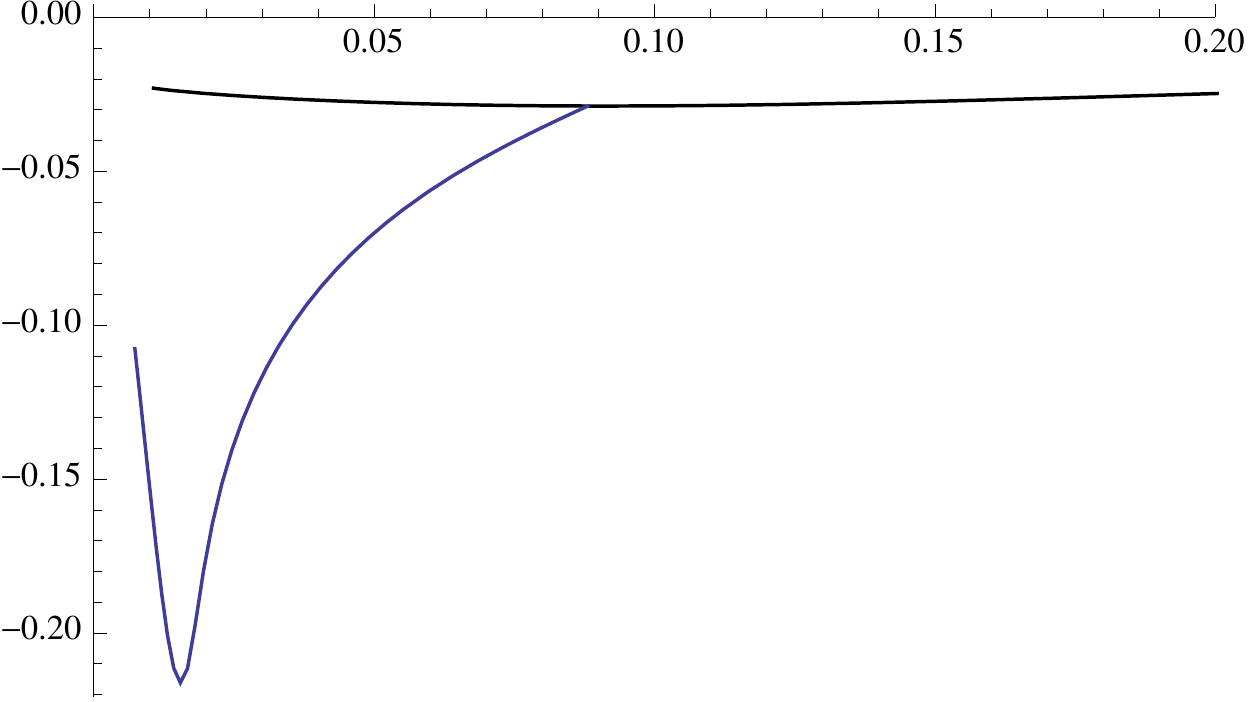}
\hskip2em
\includegraphics[width=0.4\textwidth]{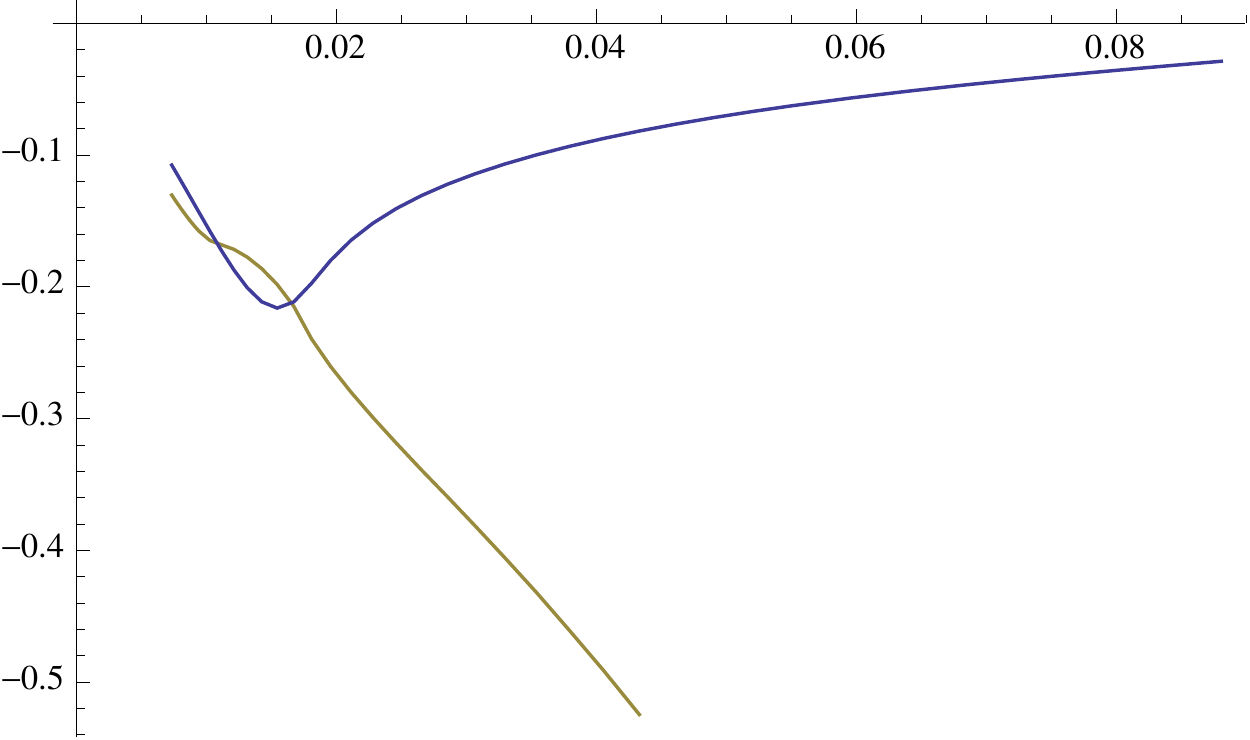}
\setlength{\unitlength}{0.1\textwidth}
\begin{picture}(0.3,0.4)(0,0)
\put(-9.1,2.){\makebox(0,0){$\Im\frac{\omega}{\mu}$}}
\put(0.2,2.3){\makebox(0,0){$\frac{T}{\mu}$}}
\end{picture}
\end{center}
\vskip-1em
\caption{Poles  in $G^R_{J^x J^x}$ closest to the real axis as a function of temperature at $\alpha/\mu = 0.3$.  The black curve corresponds to the purely imaginary Drude pole in the normal phase. For the broken phase we plot the purely imaginary pole in blue (appearing in both plots) and the (imaginary part of the) propagating pole in yellow.}\label{fig:axion_QNM} 
\end{figure}

Instead, an explanation for this feature is provided by studying the poles of  the retarded Green's function $G^R_{J^x J^x}$ directly.  In figure~\ref{fig:axion_QNM} we plot the imaginary part of the two poles closest to the axis for the broken phase as well as the Drude pole for the normal phase.  From the left-hand plot, we observe that the longest-lived purely imaginary pole in the broken phase is continuously connected to the Drude pole of the normal phase, as one would expect.  It is curious that the minimum of the normal phase curve occurs  precisely at the critical temperature. For temperatures close to $T_c/\mu$ we see from the right-hand plot that the purely imaginary mode in the broken phase is isolated, which is why the fit to a Drude form still works. However, as the temperature is lowered, a pair of propagating poles comes up from lower down in the complex plane and the Drude pole is no longer isolated. Thus the formula \eqref{2 fluid model} for the low frequency conductivity should be modified at low temperatures.  Note that this pair is not connected to the lowest propagating pair in the normal phase.

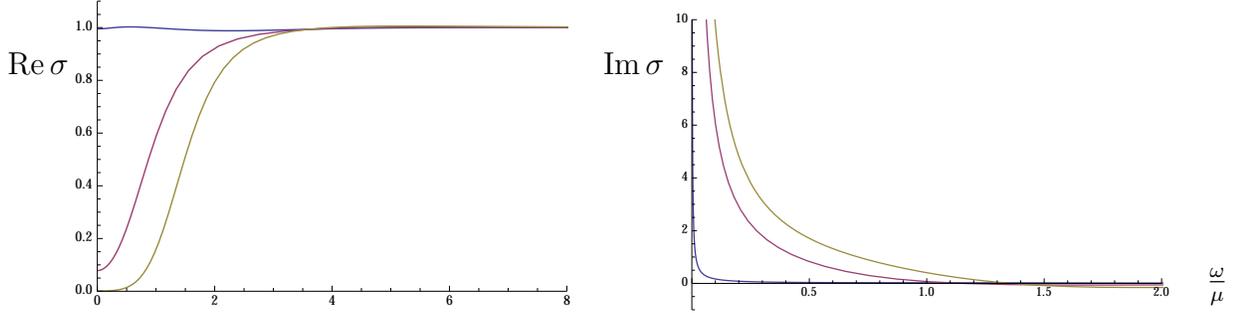
\begin{figure}[!ht]
\begin{center}
\hskip1em
\includegraphics[width=0.4\textwidth]{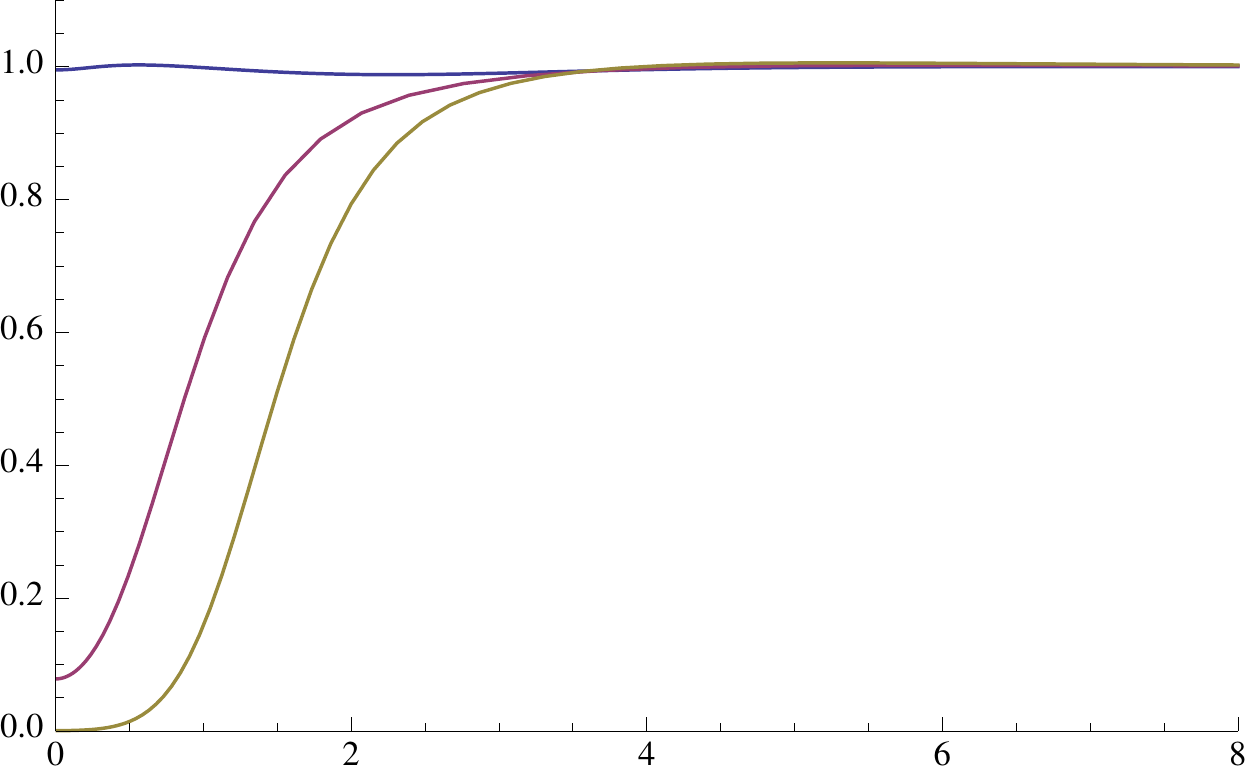}
\hskip3em
\includegraphics[width=0.4\textwidth]{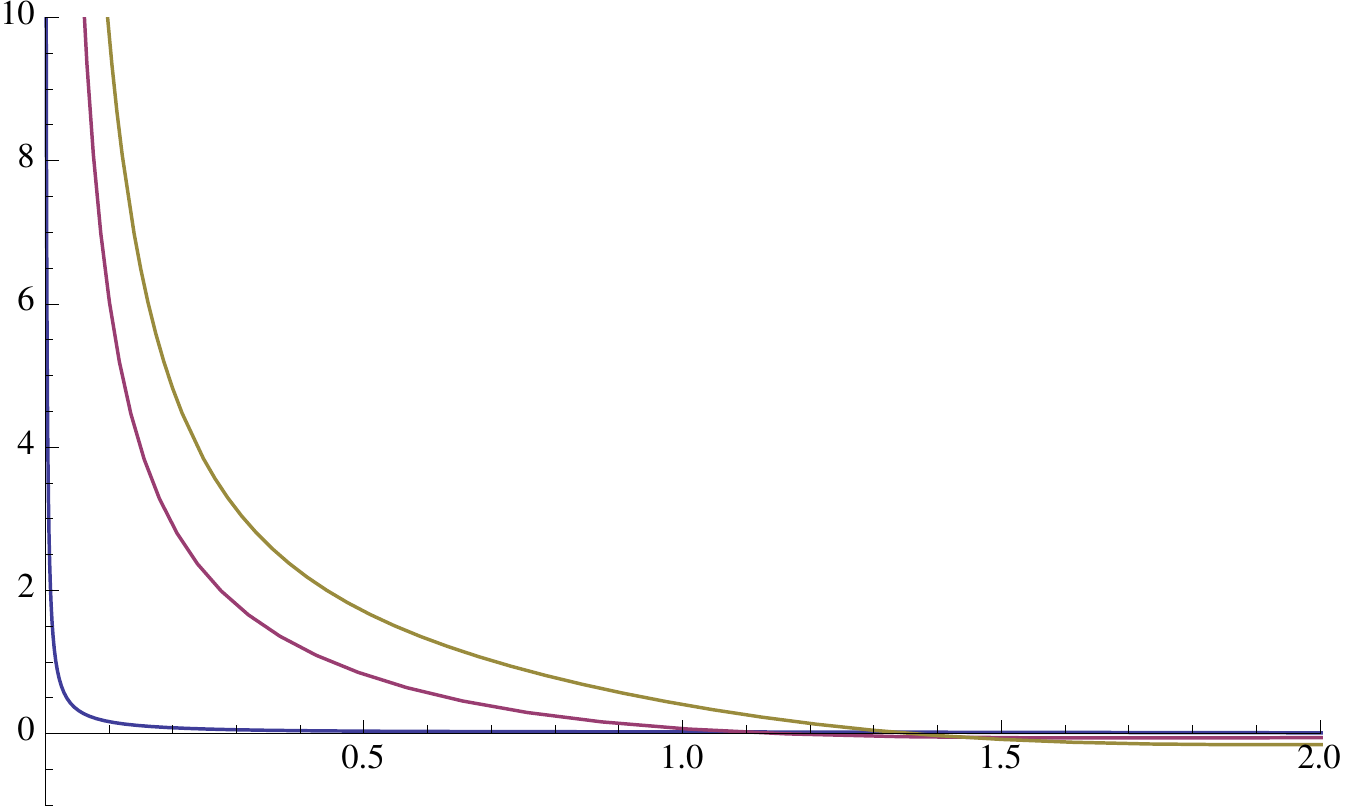}
\setlength{\unitlength}{0.1\textwidth}
\begin{picture}(0.3,0.4)(0,0)
\put(-9.3,2.){\makebox(0,0){$\Re\sigma$}}
\put(-4.5,2.){\makebox(0,0){$\Im\sigma$}}
\put(0.2,0.2){\makebox(0,0){$\frac{\omega}{\mu}$}}
\end{picture}
\end{center}
\vskip-1em
\caption{Conductivity in  the broken phase at $\alpha/\mu = 5$ as a function of frequency for  temperatures $T/\mu=0.074\lesssim T_c/\mu, 0.057, 0.029$, labelled from top to bottom in the left plot and from bottom to top in the right.}\label{fig:axion_sigma_large_alpha} 
\vskip-1em
\end{figure}

In figure~\ref{fig:axion_sigma_large_alpha} we present the optical conductivity in the broken phase at the much larger value of $\alpha/\mu = 5$.  It is clear that there is no coherent Drude-like peak, for any temperature lower than $T_c/\mu$. 

\section{Superconductivity in an axion-dilaton model}\label{sec:axion_dilaton}

In this section we  study a model that has neutral sector \cite{Donos:2013eha}
\begin{equation}
I_{\textrm{N}}= \int d^4x\sqrt{-g} \left[ R +6 - \frac{1}{4} F^2 - | \partial \Phi|^2 - m_\Phi^2|\Phi|^2   \right ]
\end{equation}
By parametrizing the complex scalar as $\Phi= \phi\, e^{i\chi}$ we can re-write this theory as an axion-dilaton model.  In comparison to the previous model we now only have a single axion, so we must build anisotropic normal phase solutions, which we have only been able to find numerically. We also have a neutral dilaton field that extends the phase space of solutions.  We fix its mass to be $m_\Phi^2=-2$.  
Allowing for these differences, we proceed in a similar manner as for the previous model and add to this action the same charged sector $I_{\textrm{C}}$ \eqref{eq:chargedaction} as before.

We consider the anisotropic ansatz 
\begin{gather}
	ds^2 = - U(r) dt^2  + \frac{dr^2}{U(r)} + e^{2V_x(r)} dx^2 + e^{2V_y(r)}dy^2 \\
	A = A_t(r) dt, \quad \Phi = e^{ikx}\phi(r),  \quad \psi = \psi(r)
\end{gather}
and study black brane solutions with temperature 
\begin{equation}
T = \frac{U'(r_+)}{4 \pi} = \frac{1}{16 \pi V_y'(r_+)} \left(12-A_t'(r_+)^2+4 \phi(r_+)^2+4 \psi(r_+)^2\right)
\end{equation}
The equations of motion admit the following asymptotic expansion near the AdS$_4$ boundary: 
\begin{align}
A_t &= \mu - \frac{\rho}{r}+ \ldots \\
\phi &= \frac{\phi_1}{r}+\frac{\phi_2}{r^2}+\ldots \\
\psi &= \frac{\psi_1}{r}+ \frac{\psi_2}{r^2}+\ldots \\
U &= r^2 -\frac{\phi_1^2+ \psi_1^2}{2} -\frac{m}{r}+\ldots\\
V_x & = \log r -\frac{\phi_1^2+\psi_1^2}{4 r^2} + \frac{V_{\mathrm{UV}}}{ r^3} + \ldots \\
V_y & = \log r -\frac{\phi_1^2+ \psi_1^2}{4 r^2} - \frac{V_{\mathrm{UV}}+2( \phi_1 \phi_2 + \psi_1 \psi_2)/3 }{ r^3} + \ldots 
\end{align}
We allow a source $\phi_1\neq 0$ for the neutral scalar. This provides an extra direction to the phase space of the dual theory (now parametrized by three dimensionless quantities $T/\mu$, $k/\mu$ and $\phi_1/\mu$), allowing for a very rich structure.  In particular, by tuning these parameters one can exhibit transitions between metallic and insulating states, as was demonstrated in a similar model in \cite{Donos:2014uba}. However, in this paper we restrict to two choices: $(k/\mu,\phi_1/\mu)=(1/2,1/\sqrt{2})$ describing a metal and $(2,2^{-3/2})$ describing an insulator.  This characterization will be examined in the following section. As before, we are interested in solutions with $\psi_2 \neq 0$ and $\psi_1 = 0$ that correspond to turning on a charged condensate in these states.  

Following \cite{Donos:2013eha}, we compute the optical conductivity by considering coupled fluctuations 
of the gauge field, metric and the dilaton field of the form
\begin{equation}
\delta g_{tx} = h_{tx}(t,r), \quad \delta A_{x} = a_{x}(t,r), \quad \delta \Phi = ie^{ikx}\varphi(t,r)
\end{equation}
Upon substitution into the equations of motion we obtain PDEs involving real fields which depend on $t$ and $r$.  
We  then let $X(t,r)=e^{-i\omega t} X(r)$ for each field $X$ and obtain the following ODEs:
\begin{align}
%
a_x''+ \left(\frac{U'}{U}-V_x'+V_y'\right) a_x'+ \left(\frac{\omega ^2-2 q^2 U \psi ^2}{U^2}-\frac{A_t'^2}{U}\right)\, a_x
-\frac{2 i k A_t' }{\omega}\, (\phi \varphi '-  \phi ' \varphi) &=0 \\
\varphi ''+ \left(\frac{U'}{U}+V_x'+V_y'\right)\varphi ' +\frac{ \left( \left(2-k^2 e^{-2 V_x}\right)U+\omega^2\right)}{U^2}\, \varphi -\frac{i k \omega   e^{-2 V_x} \phi }{U^2}\, h_{tx} &= 0 \\
%
h_{tx}'-2 V_x' h_{tx} +A_t' a_x +\frac{2 i k U}{\omega }\, (\phi  \varphi'- \phi' \varphi ) &=0
\end{align}
As in section~\ref{sec:axion}, the only difference when a condensate scalar $\psi$ is turned on is an additional 
mass term for $a_x$. 

Ingoing boundary conditions at the horizon are completely analogous to that of the axion model --- see (\ref{ingoing bc axion 1}--\ref{ingoing bc axion 3}). At the boundary we have 
\begin{align}
	\varphi &= \frac{\varphi^{(1)}}{r} + \frac{\varphi^{(2)}}{r^2} + \ldots   \\
	a_x &= a_x^{(0)} + \frac{a_x^{(1)}}{r}  + \ldots  \\
	h_{tx} &= r^2 h_{tx}^{(0)} + \ldots 
\end{align}
In terms of these coefficients, the gauge-invariant boundary condition required to calculate $\sigma(\omega)$ can be written as
\begin{equation}\label{gauge inv bc lin qlat}
	\omega \varphi^{(1)} - i k \phi_1 h_{tx}^{(0)} = 0
\end{equation}
We extract the conductivity \eqref{eq:sigmadef} from fluctuations that satisfy ingoing boundary conditions at the horizon and \eqref{gauge inv bc lin qlat} 
in the UV.

\subsection{Normal phase}

\begin{figure}[!ht]
\begin{center}
\includegraphics[width=0.4\textwidth]{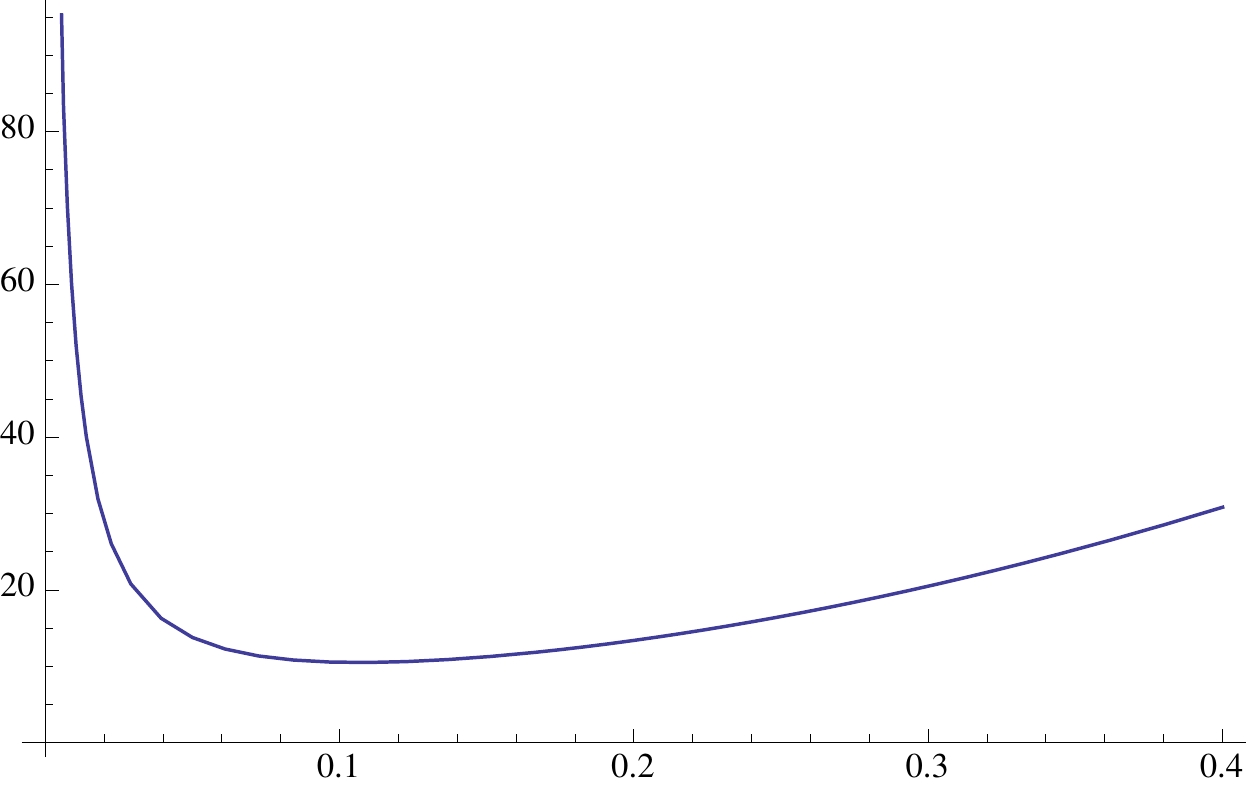}
\hskip2em
\includegraphics[width=0.4\textwidth]{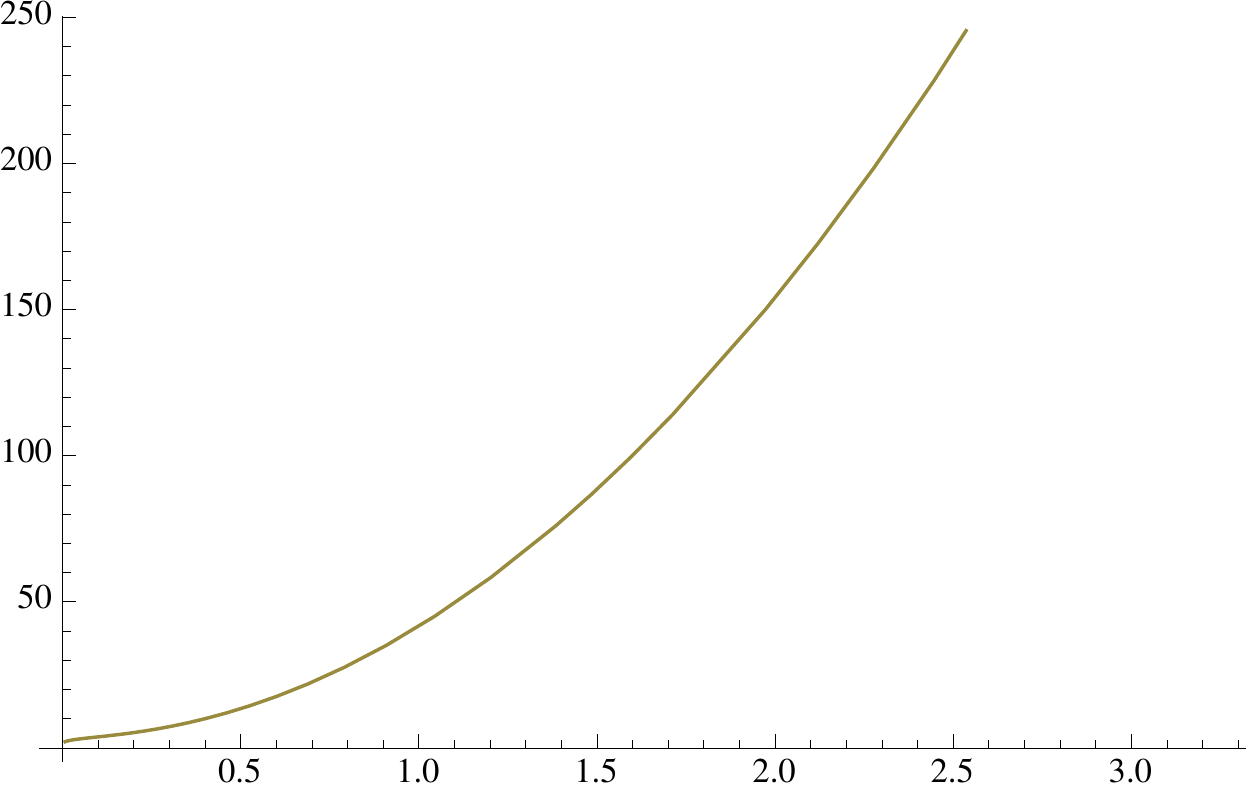}
\setlength{\unitlength}{0.1\textwidth}
\begin{picture}(0.3,0.4)(0,0)
\put(-9,2.){\makebox(0,0){$\sigma_{\mathrm{DC}}$}}
\put(0.2,0.2){\makebox(0,0){$\frac{T}{\mu}$}}
\end{picture}
\end{center}
\vskip-1em
\caption{DC conductivity as a function of temperature for the metal  (left) and the insulator (right).}\label{fig:axion_dilaton_sigma_DC} 
\vskip-1em
\end{figure}
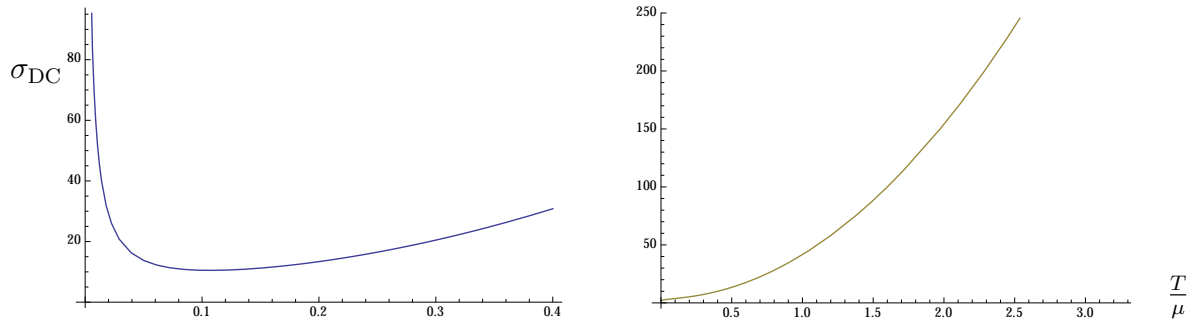

Normal phase solutions $\psi=0$ were studied in  \cite{Donos:2013eha}.  A formula for the DC conductivity in terms of the solution data was derived  for a general class of axion-dilaton models in \cite{Donos:2014uba,Gouteraux:2014hca}.  In this particular model it reduces to
\begin{equation}\label{DG sigma DC}
\sigma_{\textrm{DC}} = e^{-V_x(r_+)+V_y(r_+)} + \frac{e^{V_x(r_+)+V_y(r_+)} A_t'(r_+)^2}{2 k^2  \phi(r_+)^2}  \end{equation}
An insulator has vanishing DC conductivity at zero temperature.  In figure~\ref{fig:axion_dilaton_sigma_DC} we plot the DC conductivity as a function of temperature for our two points in phase space, confirming their metallic or insulating character.

It is worth noting that the DC resistivity as a function of temperature in the metallic state is not monotonic but possesses a peak at finite temperature. This feature is also present in the model of \cite{Taylor:2014tka} and it is observed in some experimental heavy fermion setups.\footnote{We thank Marika Taylor for sharing this unpublished result with us.}

For the metal, the low frequency behaviour of the conductivity is well-approximated by a Drude peak 
even at very low temperatures \cite{Donos:2013eha}. In the insulating state, on the other hand,  Drude behaviour ceases to be 
a good approximation.  In either case, equation \eqref{DG sigma DC} gives the correct value of the DC conductivity. 
A full mapping of the phase diagram is lacking, and for now we will content ourselves with exploring a representative 
of the metallic and a representative of the insulating phases. We expect the Drude/non-Drude behaviour of the optical conductivity 
to be related to the structure of the QNM in analogy with \cite{Davison:2014lua}, but we leave this interesting question for future work.

\subsection{Broken phase}\label{sec:axion_dilaton_broken}

\begin{figure}[!ht]
\begin{center}
\includegraphics[width=0.5\textwidth]{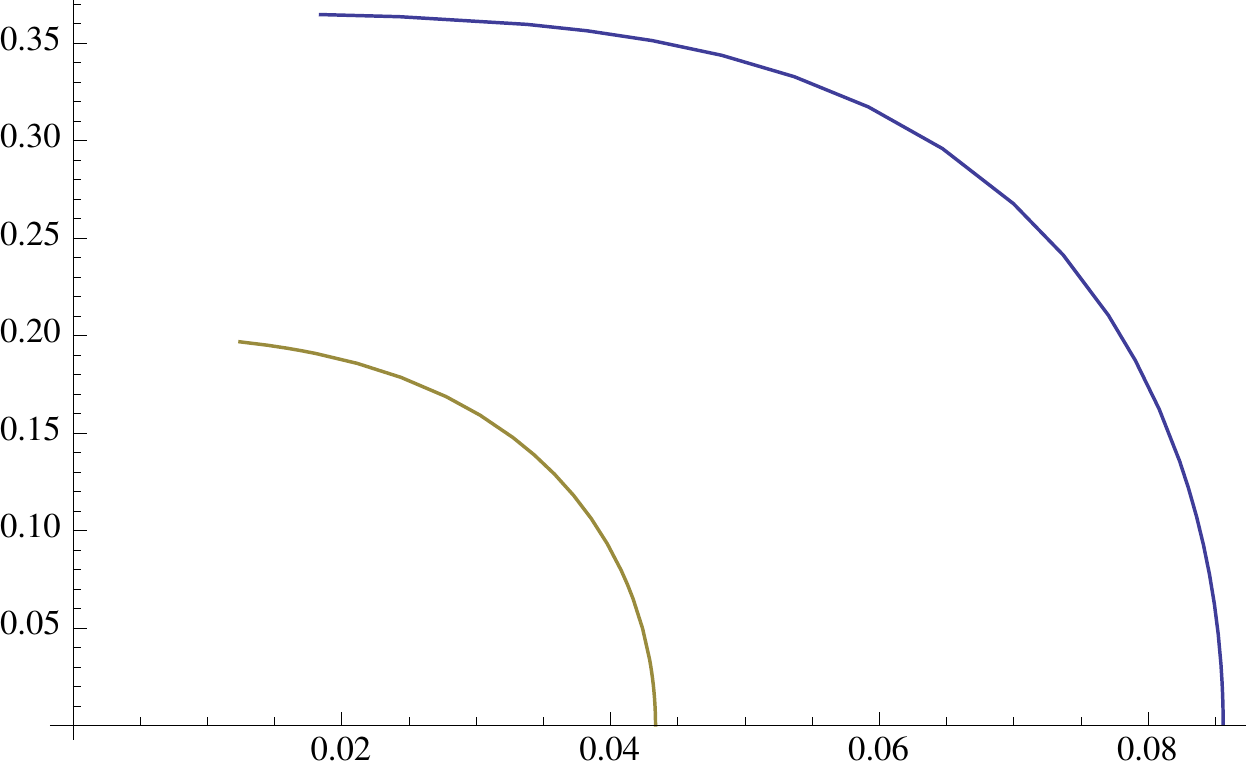}
\setlength{\unitlength}{0.1\textwidth}
\begin{picture}(0.3,0.4)(0,0)
\put(-5.7,2.5){\makebox(0,0){$\frac{\psi_2}{\mu^2}$}}
\put(0.2,0.2){\makebox(0,0){$\frac{T}{\mu}$}}
\end{picture}
\end{center}
\vskip-1em
\caption{Condensate as a function of temperature for the broken metal (blue) and the broken insulator  (yellow).}\label{fig:axion_dilaton_condensate} 
\vskip-1em
\end{figure}
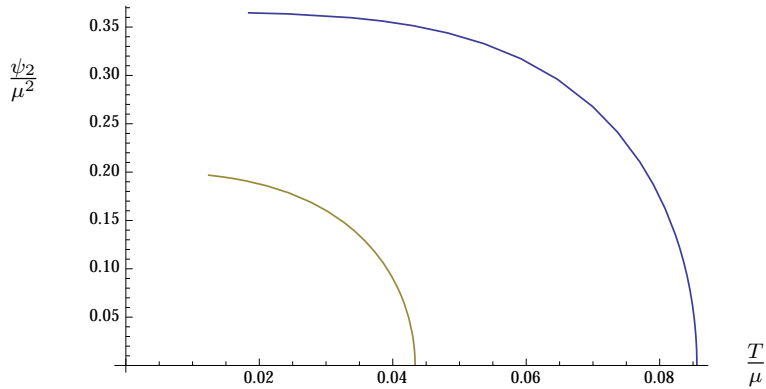

Next we study solutions with the charged scalar turned on.  We construct these numerically using a shooting method.  For the same two choices of $(k/\mu,\phi_1/\mu)$ we find black brane solutions with charged scalar hair that exist below some critical $T_c/\mu$, corresponding to the broken phases of these two states.  We will refer to the new phases as the broken metal and the broken insulator.  In figure~\ref{fig:axion_dilaton_condensate} we show condensate curves for the two $(k/\mu,\phi_1/\mu)$.  

\begin{figure}[!ht]
\begin{center}
\includegraphics[width=0.4\textwidth]{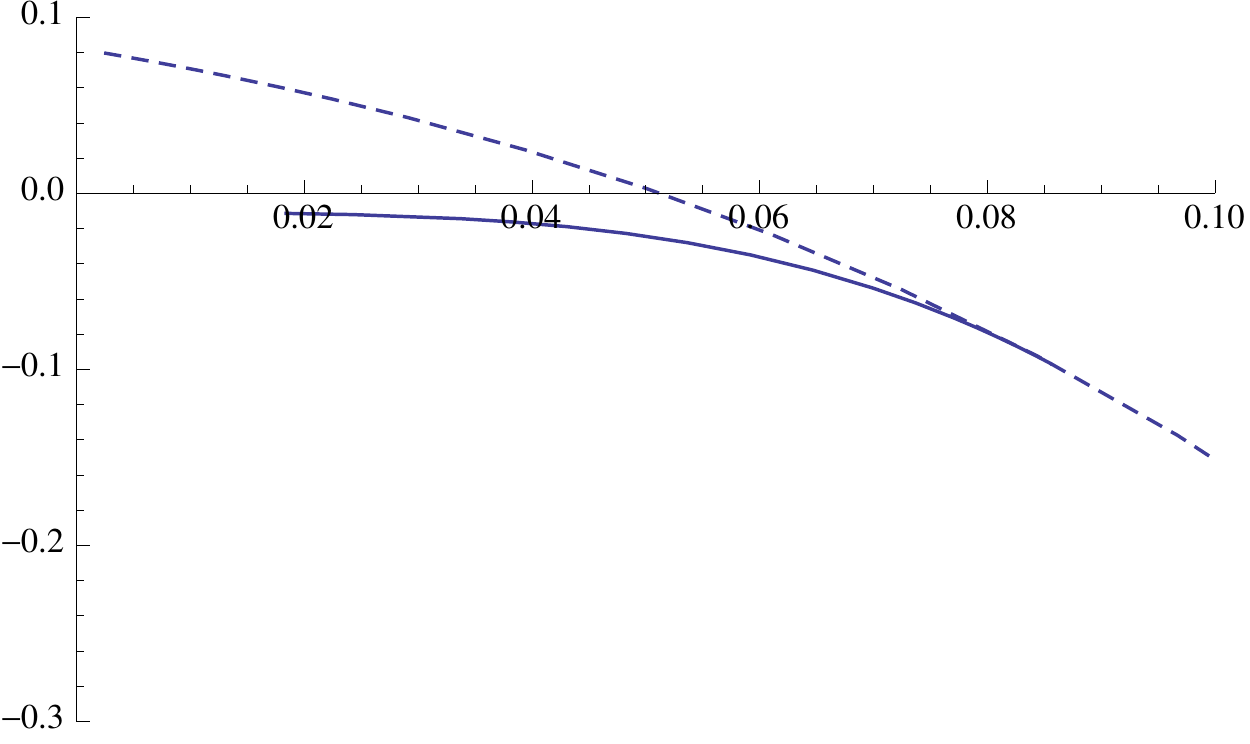}
\hskip2em
\includegraphics[width=0.4\textwidth]{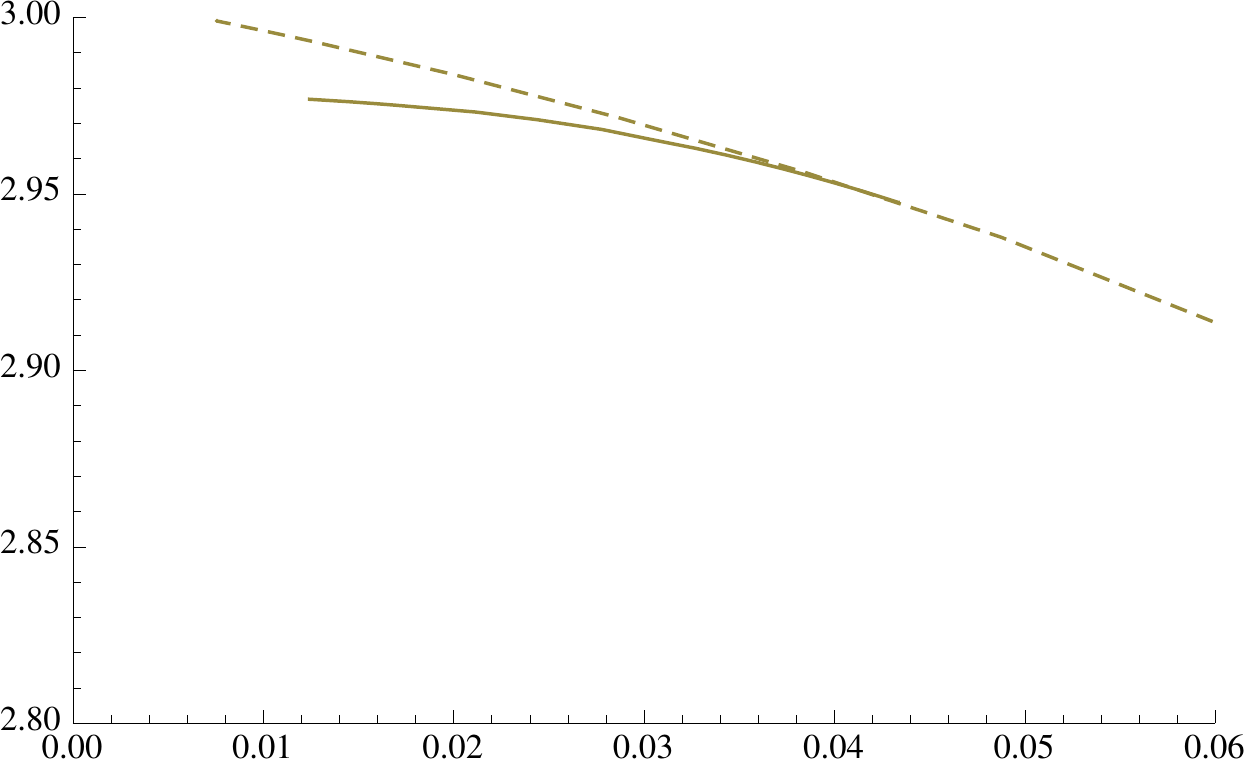}
\setlength{\unitlength}{0.1\textwidth}
\begin{picture}(0.3,0.4)(0,0)
\put(-8.9,2.){\makebox(0,0){$\frac{w}{\mu^3}$}}
\put(0.2,0.2){\makebox(0,0){$\frac{T}{\mu}$}}
\end{picture}
\end{center}
\vskip-1em
\caption{Free energy density as a function of temperature for the broken phase (solid lines) and the normal phase (dashed lines) for the metal  (left) and the insulator  (right).}\label{fig:axion_dilaton_free_energy} 
\vskip-0.5em
\end{figure}
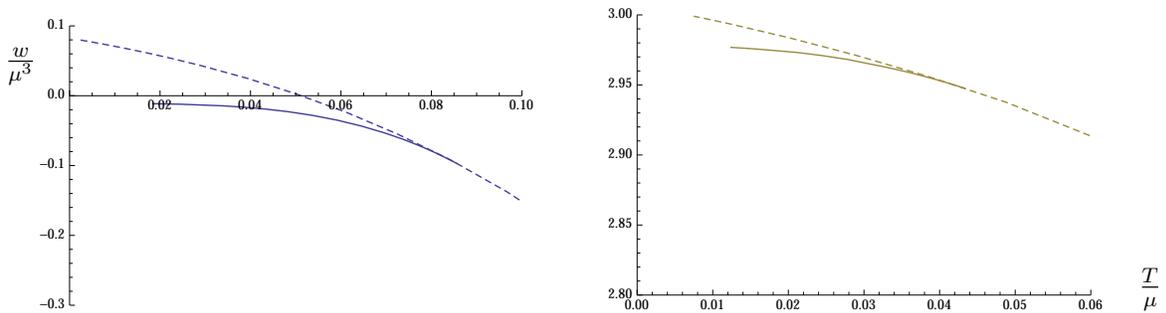

The method for computing the free energy density is similar to that for the axion model. Evaluating the Euclidean on-shell action
yields
\begin{equation}
I_E  = \beta V_2 \lim_{r\to\infty} 2\sqrt{g}\, U\, V_y' 
\end{equation}
The renormalised  action is given by
\begin{equation}
I_E^{\textrm{ren}} = I_E + \int d^3x \sqrt{h}\left( -2 \mathcal{K}+4+ |\Phi|^2+|\psi|^2\right) \\
\end{equation}
and we eventually find
\begin{equation}
w = -m + 6V_{\mathrm{UV}} + 2 \phi_1 \phi_2
\end{equation}
This formula is valid for both the  normal and broken phases.   In figure~\ref{fig:axion_dilaton_free_energy} we demonstrate that the broken phase has lower free energy density than the normal phase at the same $(k/\mu,\phi_1/\mu)$ and thus is thermodynamically preferred.

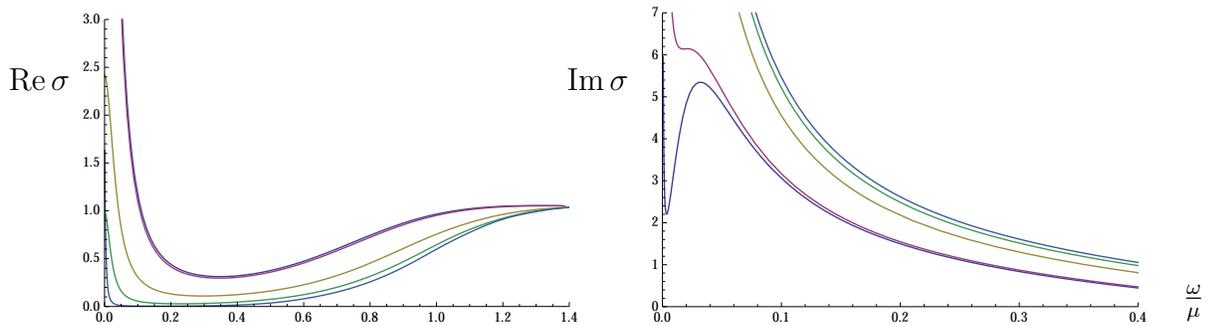
\begin{figure}[!ht]
\vskip1em
\begin{center}
\hskip1em
\includegraphics[width=0.4\textwidth]{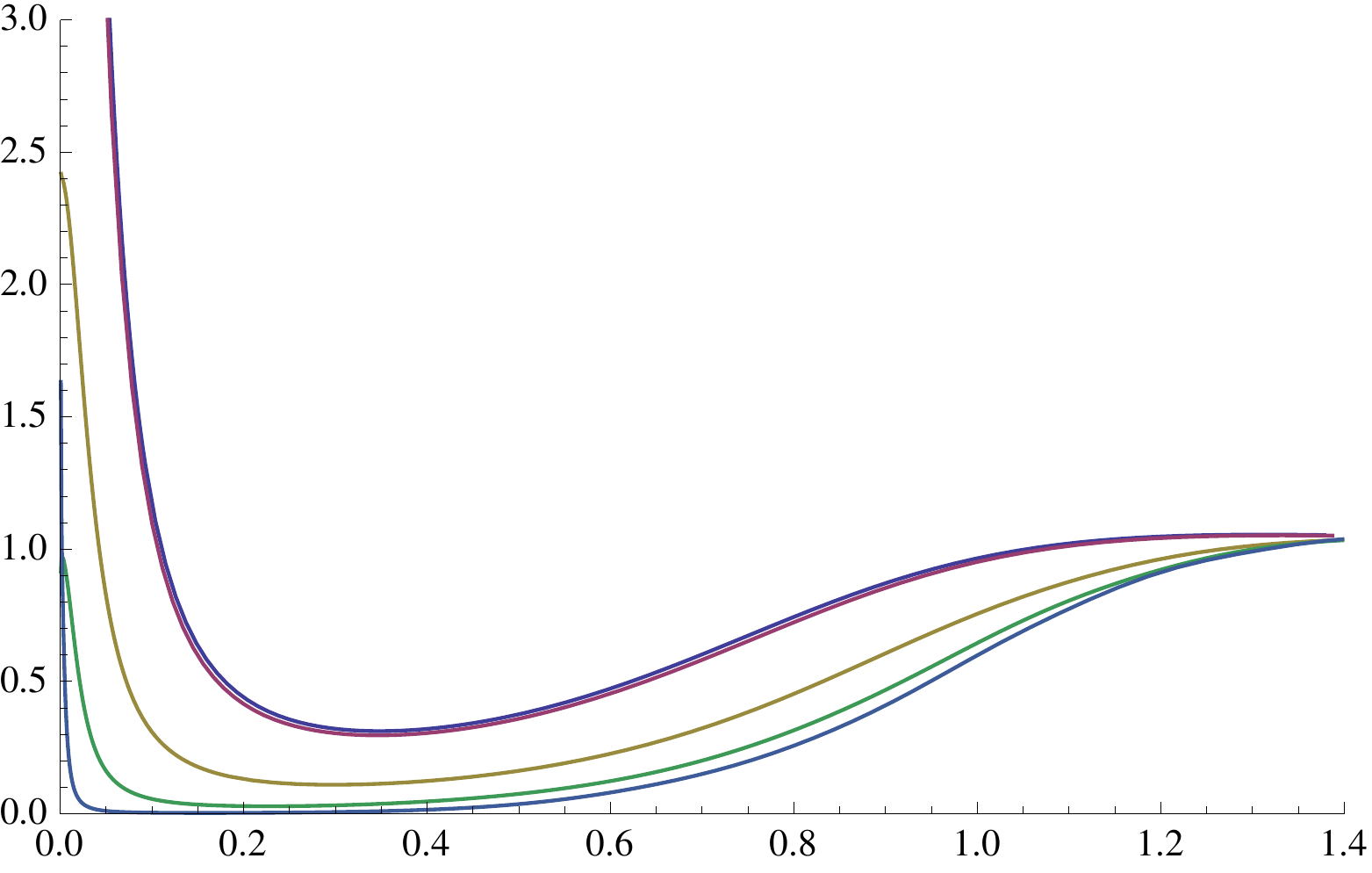}
\hskip2em
\includegraphics[width=0.4\textwidth]{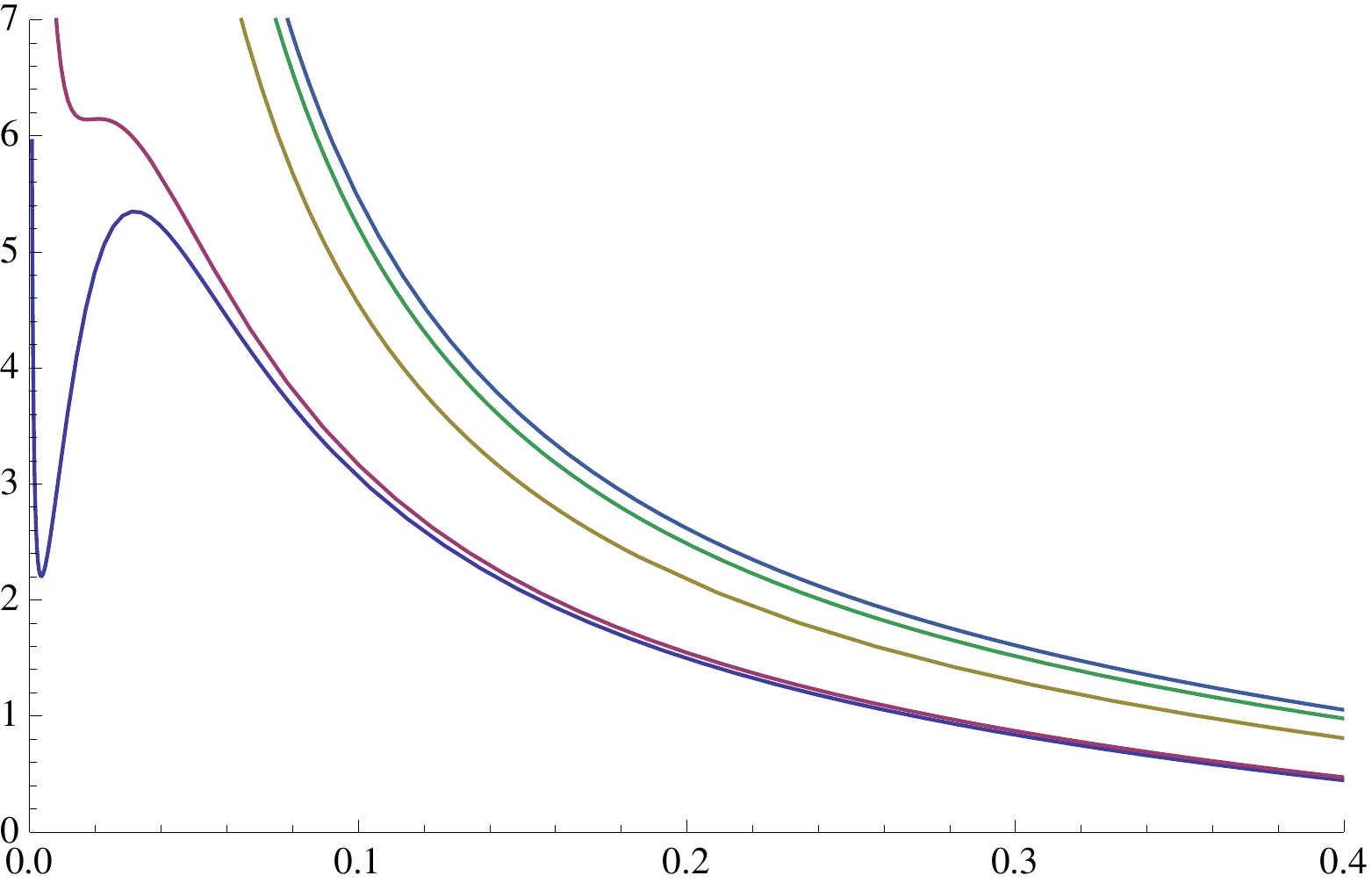}
\setlength{\unitlength}{0.1\textwidth}
\begin{picture}(0.3,0.4)(0,0)
\put(-9.1,2.){\makebox(0,0){$\Re\sigma$}}
\put(-4.6,2.){\makebox(0,0){$\Im\sigma$}}
\put(0.2,0.2){\makebox(0,0){$\frac{\omega}{\mu}$}}
\end{picture}
\end{center}
\vskip-1em
\caption{Conductivity in  the broken metal as a function of frequency for various temperatures. For the plot of the real part (left), temperature decreases from $T_c/\mu = 0.085$ to $T/\mu = 0.035$, from top to bottom in the left plot and from bottom to top in the right .}\label{fig:metal_sigma} 
\end{figure}

\begin{figure}[!ht]
\begin{center}
\includegraphics[width=0.4\textwidth]{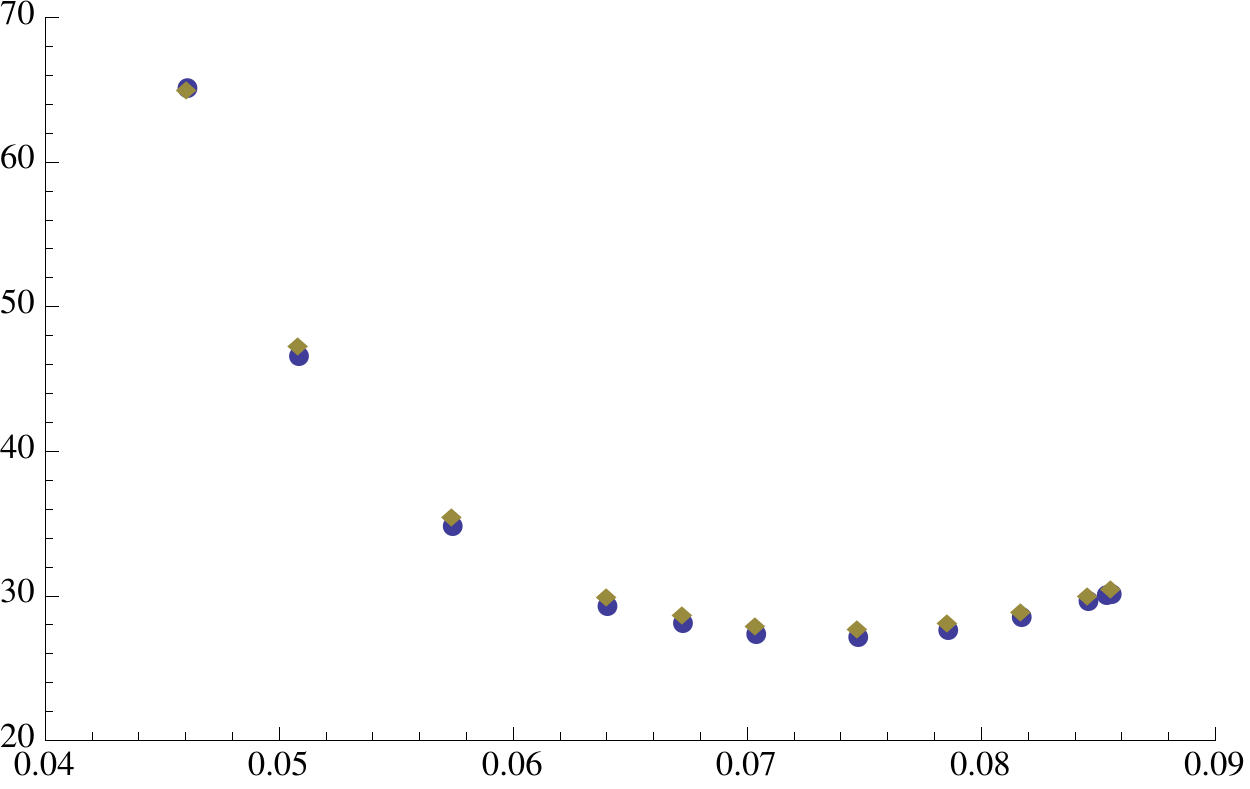}
\hskip2em
\includegraphics[width=0.4\textwidth]{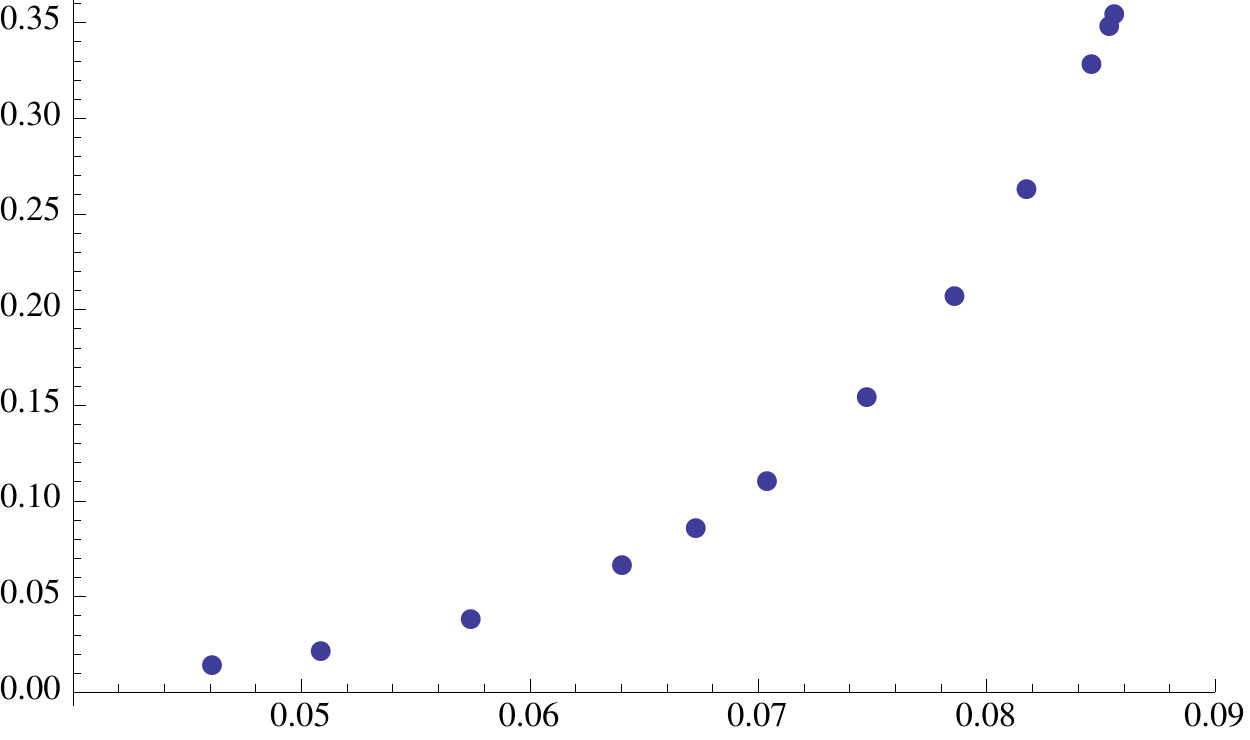}
\vskip1em
\hskip1em
\includegraphics[width=0.4\textwidth]{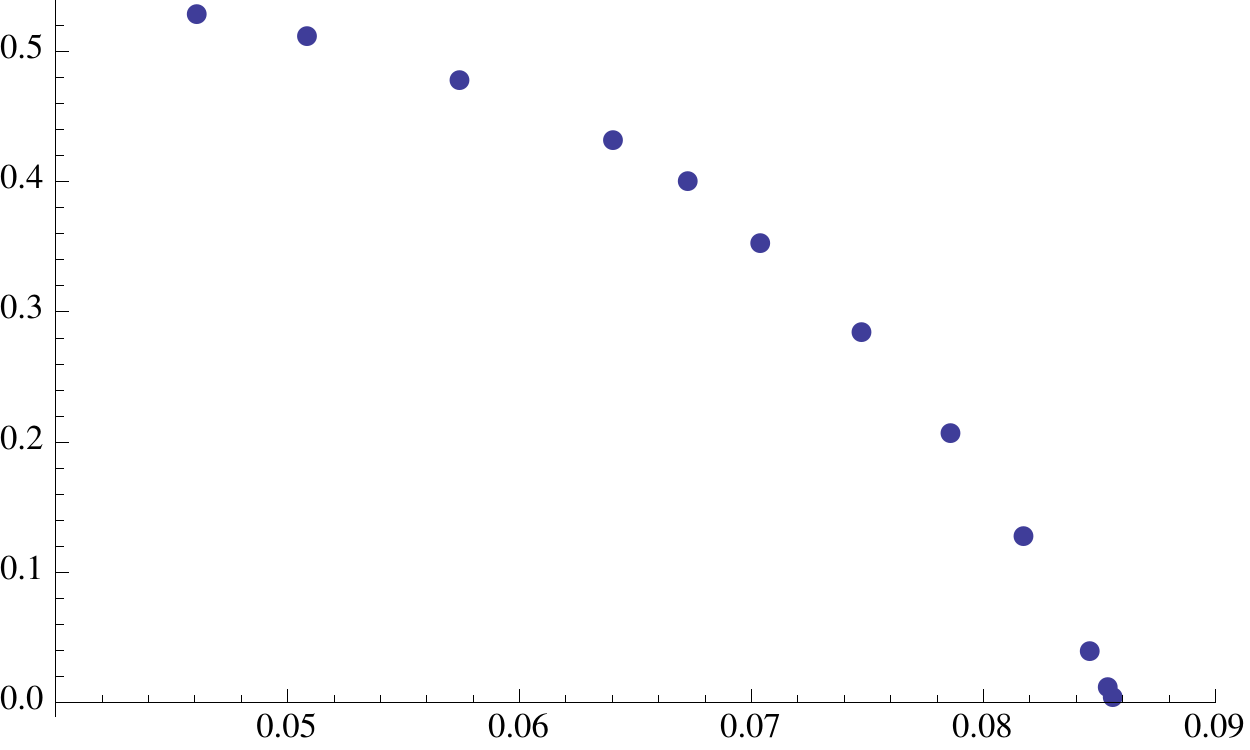}
\setlength{\unitlength}{0.1\textwidth}
\begin{picture}(0.3,0.4)(0,0)
\put(-6.8,4.8){\makebox(0,0){$\tau\mu$}}
\put(-2.1,4.8){\makebox(0,0){$\frac{K_n}{\mu}$}}
\put(2.7,2.85){\makebox(0,0){$\frac{T}{\mu}$}}
\put(-4.6,2.){\makebox(0,0){$\frac{K_s}{\mu}$}}
\put(0.2,0.15){\makebox(0,0){$\frac{T}{\mu}$}}
\end{picture}
\end{center}
\vskip-1em
\caption{Parameters of the two-fluid model \eqref{2 fluid model} as a function of temperature for the broken metal. In the plot of $\tau\mu$ versus $T/\mu$, the blue data is obtained from the fits and the yellow data comes from the lowest purely-dissipative QNM. We find excellent agreement between these two independent procedures.}\label{fig:metal_fits}
\end{figure}
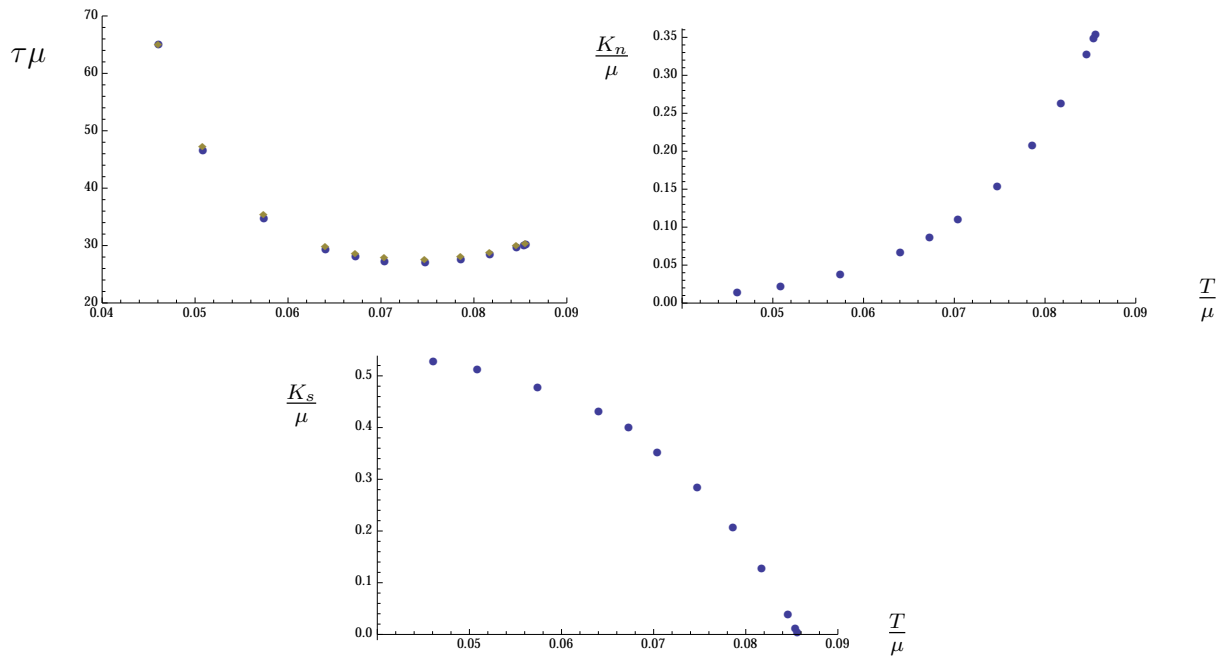

We present our results for the optical conductivity in the broken phase of the metal in figure~\ref{fig:metal_sigma}. 
In the range of temperatures we have considered, the low frequency region of the conductivity can be well-approximated by 
the two-fluid model with Drude normal component \eqref{2 fluid model}. 
More specifically, for temperatures in the range $ 0.78 \, T_c/\mu <  T/\mu  < T_c/\mu$, the value of $\tau\mu$ obtained from the 
fit to the real part of the conductivity differs by no more than  $5 \%$ from the value obtained fitting the imaginary part, and by no more
than $2 \%$ from the value obtained from the characteristic time given by the lowest (in this case, purely dissipative) QNM. For lower temperatures, the pole in the imaginary part of the conductivity overwhelms the normal component, which makes it numerically harder to perform the fit to the imaginary part. However, the value of $\tau\mu$ given by the fit to the real part still differs with the QNM value 
by only $2 \%$. For all the fits in the broken metallic case, the residuals do not exceed $10^{-3}$. 
We plot the parameters of the fit in figure \ref{fig:metal_fits}.

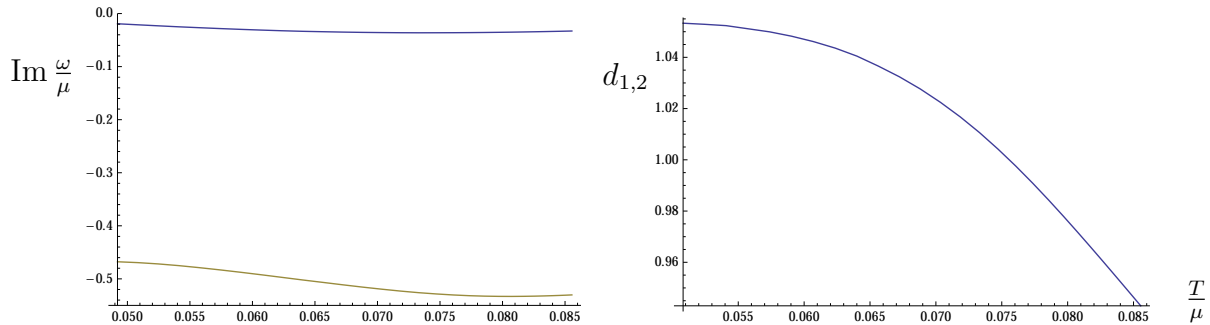
\begin{figure}[!ht]
\begin{center}
\hskip1em
\includegraphics[width=0.4\textwidth]{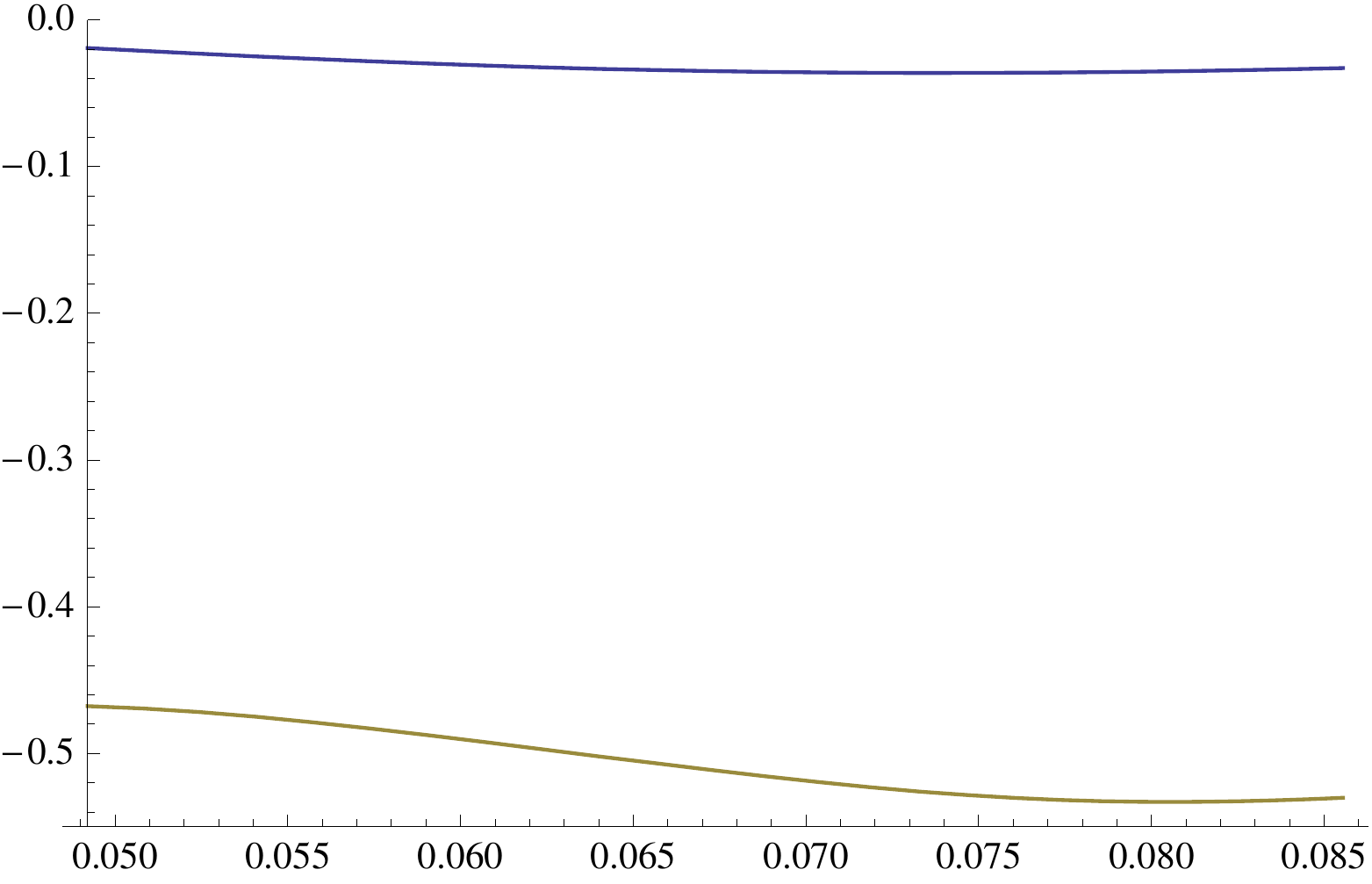}
\hskip2em
\includegraphics[width=0.4\textwidth]{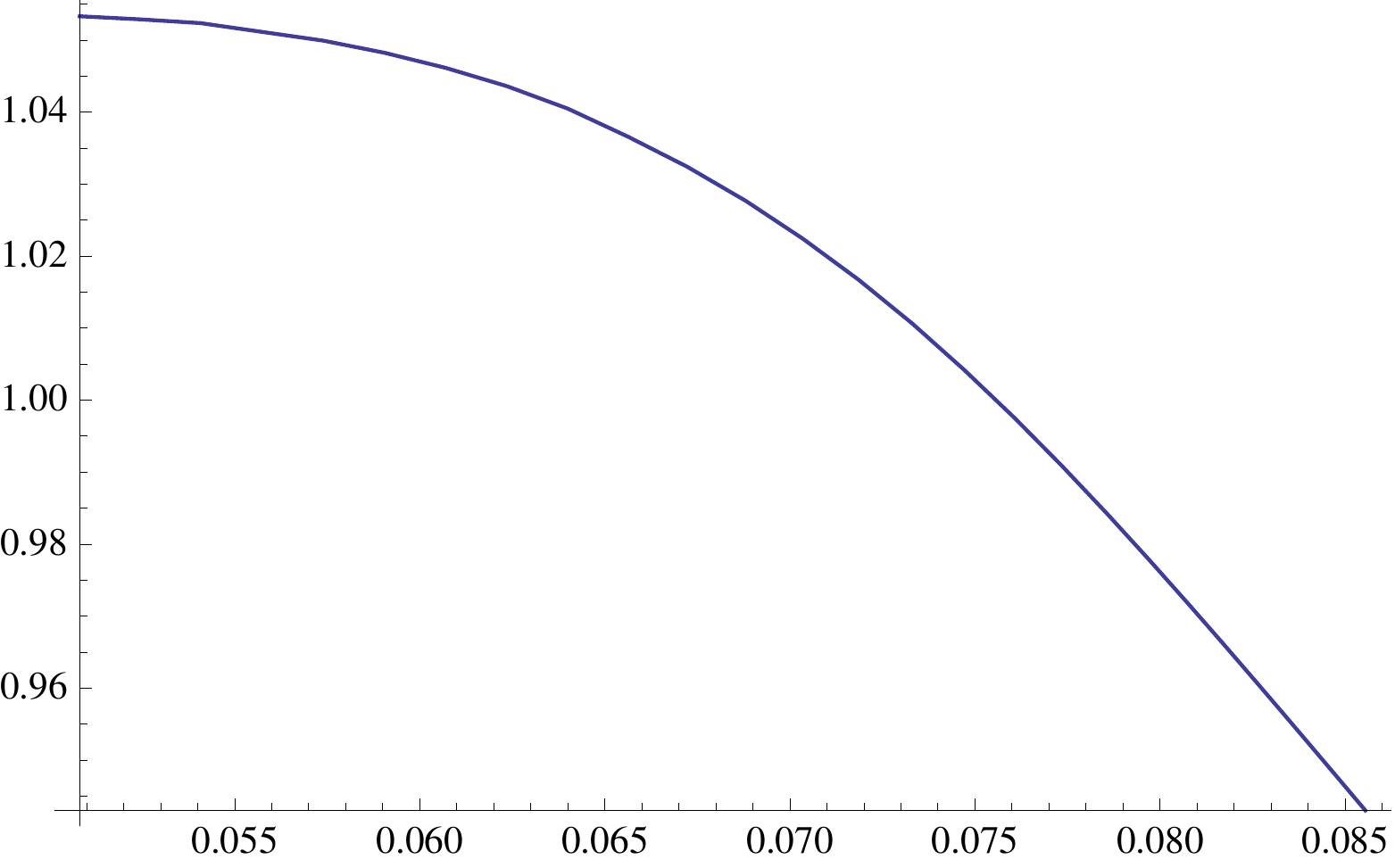}
\setlength{\unitlength}{0.1\textwidth}
\begin{picture}(0.3,0.4)(0,0)
\put(-9.1,2.){\makebox(0,0){$\Im\frac{\omega}{\mu}$}}
\put(-4.4,2.){\makebox(0,0){$d_{1,2}$}}
\put(0.2,0.2){\makebox(0,0){$\frac{T}{\mu}$}}
\end{picture}
\end{center}
\vskip-1em
\caption{Poles  in $G^R_{J^x J^x}$ closest to the real axis as a function of temperature for the broken metal.  Left: We plot the purely-imaginary pole in blue and the (imaginary part of the) propagating pole in yellow. Right: The absolute value of the difference between these two poles.}\label{fig:metal_QNM} 
\end{figure}

The structure of the low frequency region can be qualitatively understood in terms of the QNM spectrum --- 
see figure~\ref{fig:metal_QNM}. We find that 
the lowest QNM is purely dissipative and well-separated from the other excitations, which, following the arguments 
of \cite{Davison:2014lua}, accounts for the Drude behaviour. Moreover, the fact that the characteristic time obtained from the QNM calculation is in excellent agreement with the one obtained from the fits provides quantitative evidence in favour of this connection,
in addition to a non-trivial cross check of our numerical procedure.  

\begin{figure}[!ht]
\begin{center}
\includegraphics[width=0.5\textwidth]{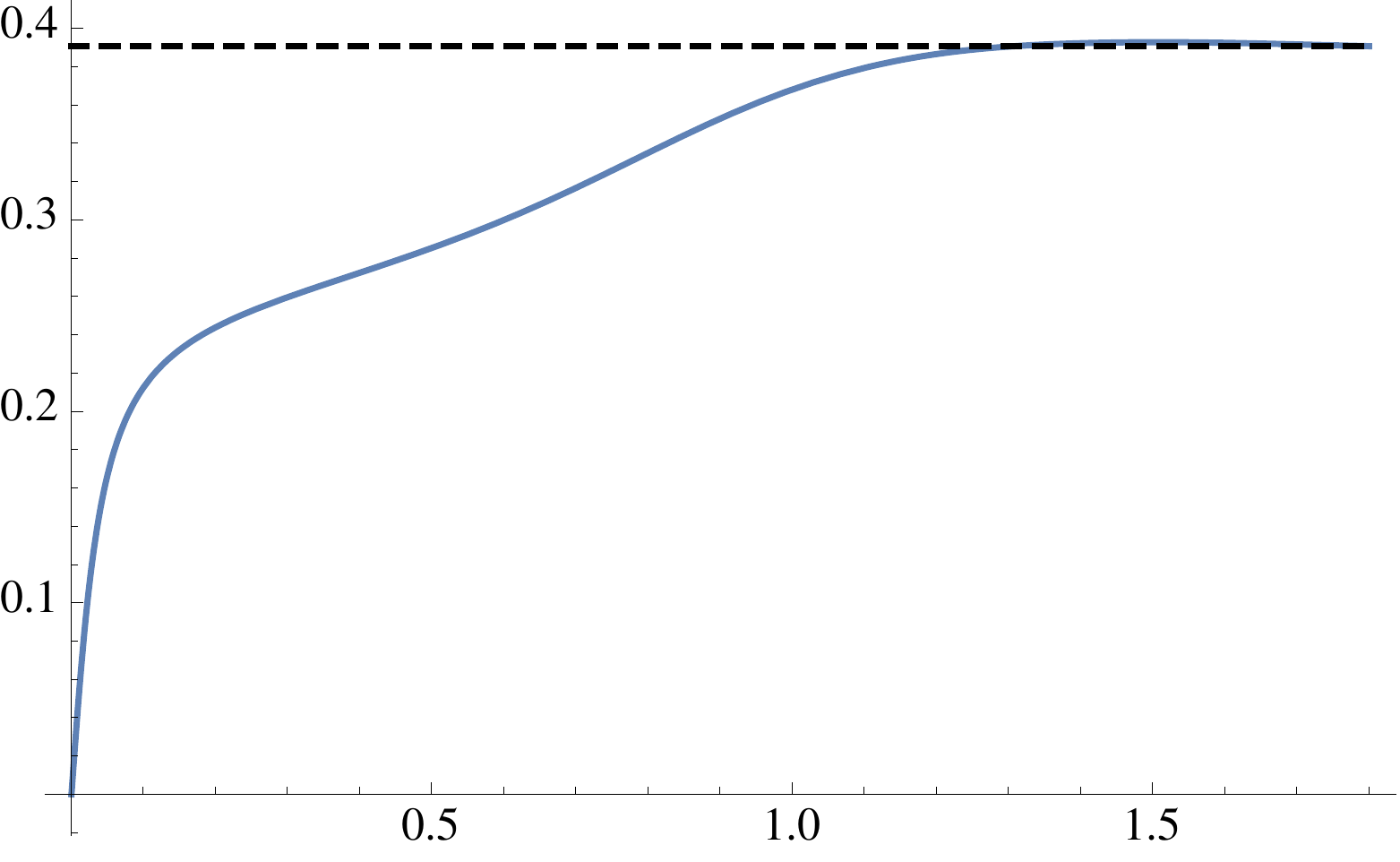}
\setlength{\unitlength}{0.1\textwidth}
\begin{picture}(0.3,0.4)(0,0)
\put(-5.7,2.5){\makebox(0,0){$F$}}
\put(0.2,0.2){\makebox(0,0){$\frac{\omega}{\mu}$}}
\end{picture}
\end{center}
\vskip-1em
\caption{Numerical check of the FGT sum rule for the metal where we choose a broken phase at
 $T/\mu =0.79\, T_c/\mu$. The dashed line is the value of $K_s/\mu = 0.39$ extracted from a fit of $\Im\sigma_s$.}
\label{fig:metal_FGT}
\end{figure}
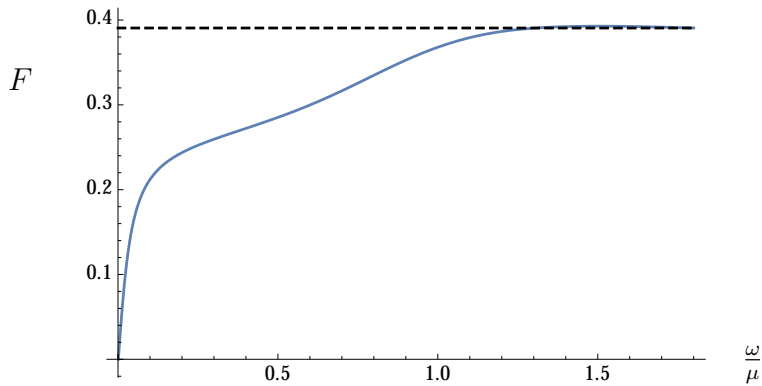

It is interesting to note that a pseudo gap forms as we lower the temperature, by which we mean that $\Re \sigma$
is very small for a range of frequencies but then the Drude peak reappears at lower frequencies. This feature was also observed in 
\cite{Koga:2014hwa}, but not in the axion model as discussed in section~\ref{sec:brokenaxion}.  We have also checked numerically that the FGT sum rule holds for  the broken metal --- see figure \ref{fig:metal_FGT}.
Once again,  we need to go to high frequencies to obtain agreement.

We present our results for the optical conductivity in the broken phase of the insulator in figure~\ref{fig:insulator_sigma}. 
Once again, we find a superconducting pole in the imaginary part of $\sigma(\omega)$. However, in contrast to 
the broken metal, the two-fluid model \eqref{2 fluid model} does not give a good description for small $\omega$. Close to $T_c$, 
the curves look qualitatively of the form \eqref{2 fluid model} but we find that the values obtained from 
fitting to the real and imaginary parts differ by more than $10 \%$. As we lower the temperature, the departure from 
\eqref{2 fluid model} is even more explicit since $\sigma(\omega)$ clearly has a qualitatively different structure. 
We argue that this is connected to the fact that, among the lowest-lying QNM, there is a purely-dissipative mode which is close
in the complex plane to a propagating mode --- see figure \ref{fig:insulator_QNM}. 
\begin{figure}[!ht]
\begin{center}
\hskip1em
\includegraphics[width=0.4\textwidth]{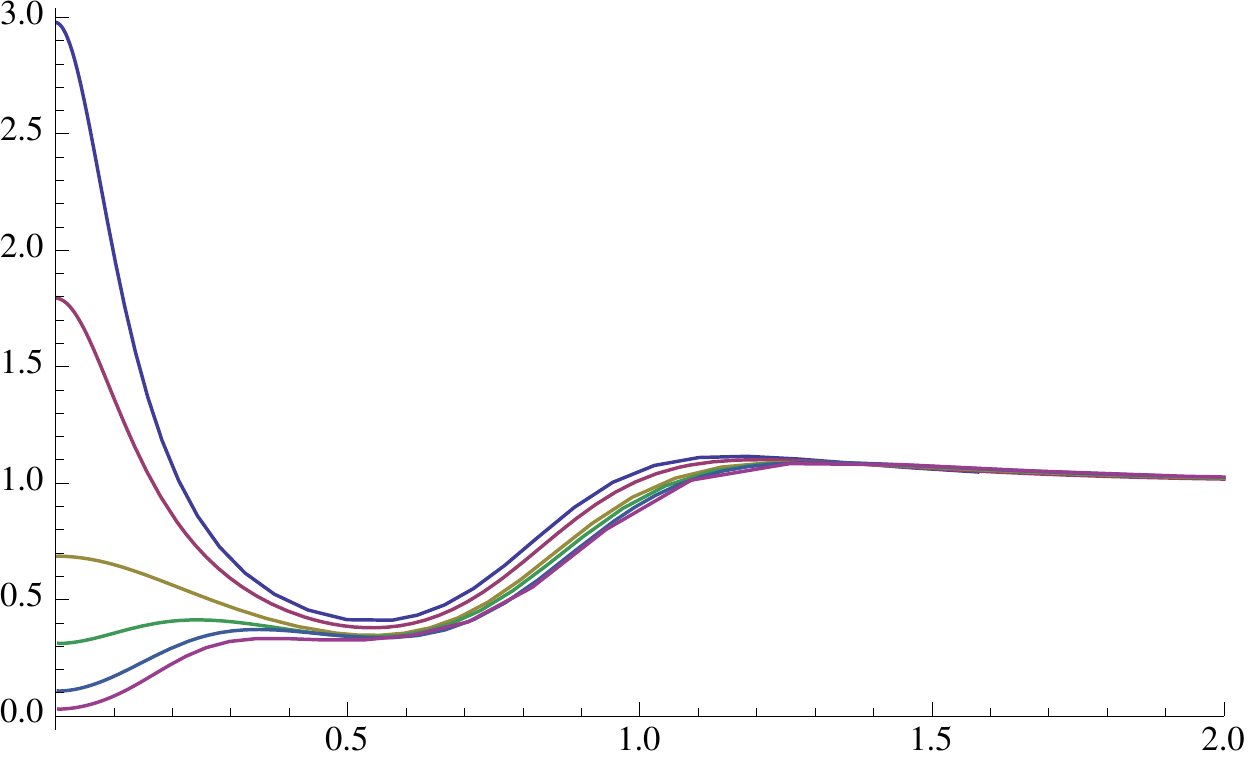}
\hskip2em
\includegraphics[width=0.4\textwidth]{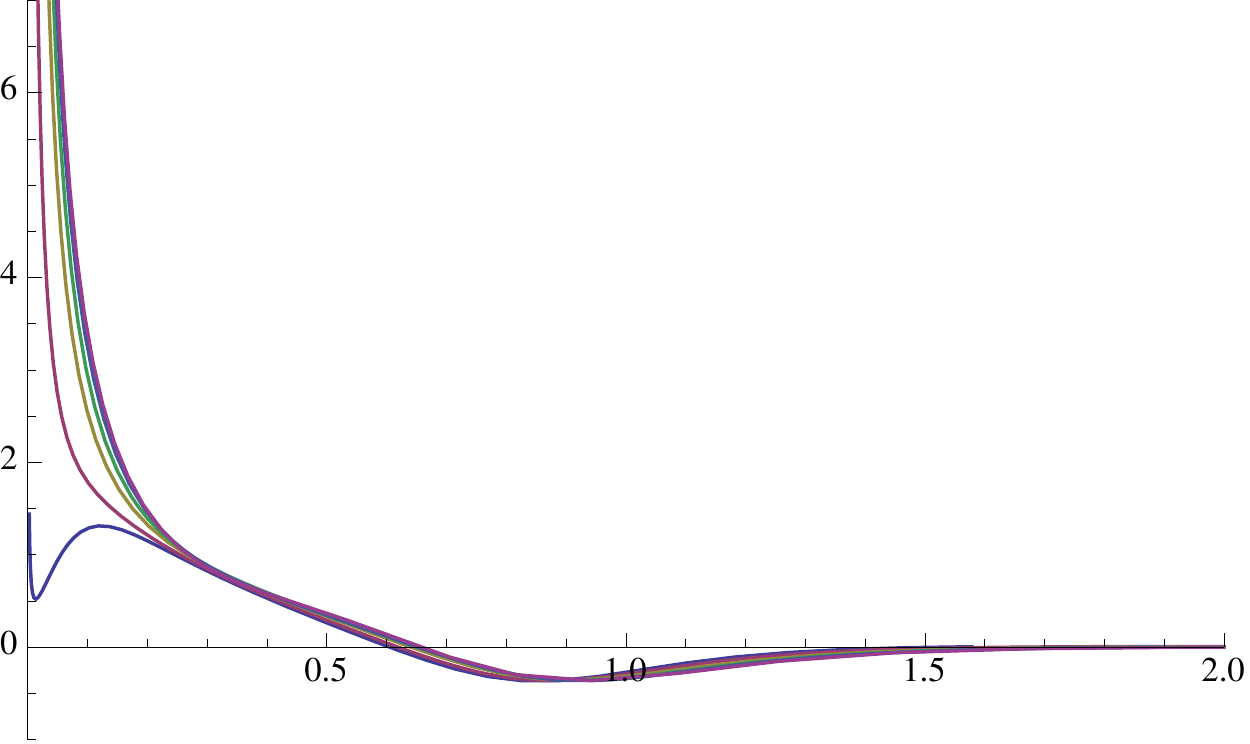}
\setlength{\unitlength}{0.1\textwidth}
\begin{picture}(0.3,0.4)(0,0)
\put(-9.1,2.){\makebox(0,0){$\Re\sigma$}}
\put(-4.6,2.){\makebox(0,0){$\Im\sigma$}}
\put(0.2,0.2){\makebox(0,0){$\frac{\omega}{\mu}$}}
\end{picture}
\end{center}
\vskip-1em
\caption{Conductivity in  the broken insulator as a function of frequency for various temperatures. 
The temperature decreases from $T_c/\mu = 0.043$ down to $T/\mu = 0.012$, from top to bottom in the left plot and from bottom to top in the right.} \label{fig:insulator_sigma} 
\end{figure}
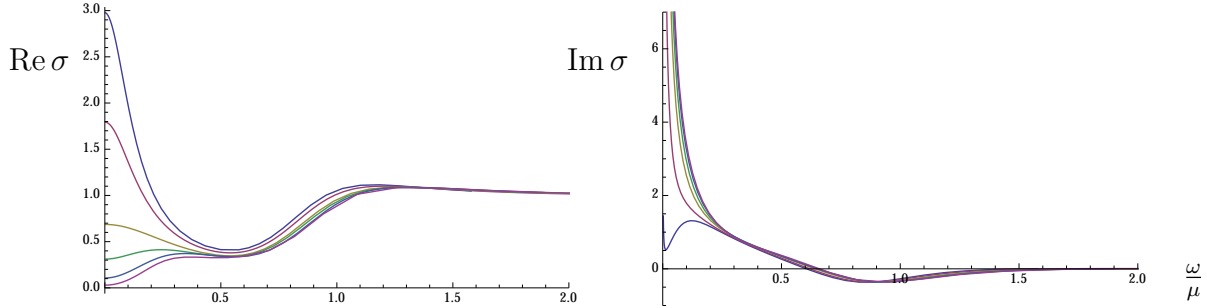 
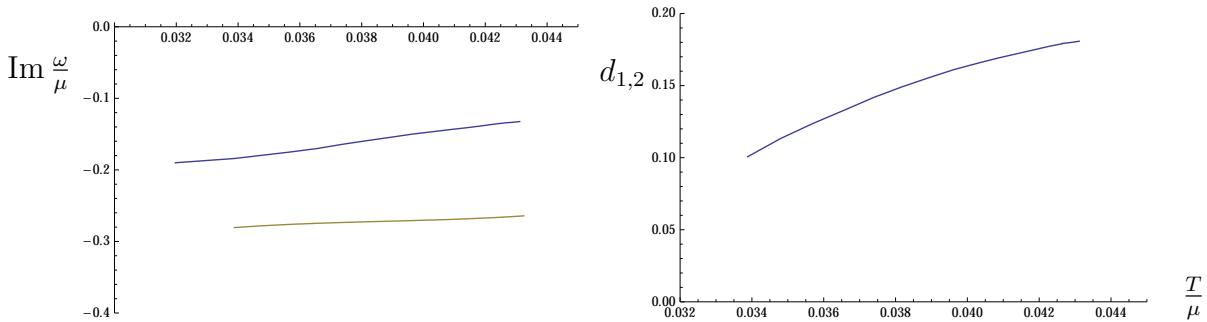
\begin{figure}[!ht]
\begin{center}
\hskip1em
\includegraphics[width=0.4\textwidth]{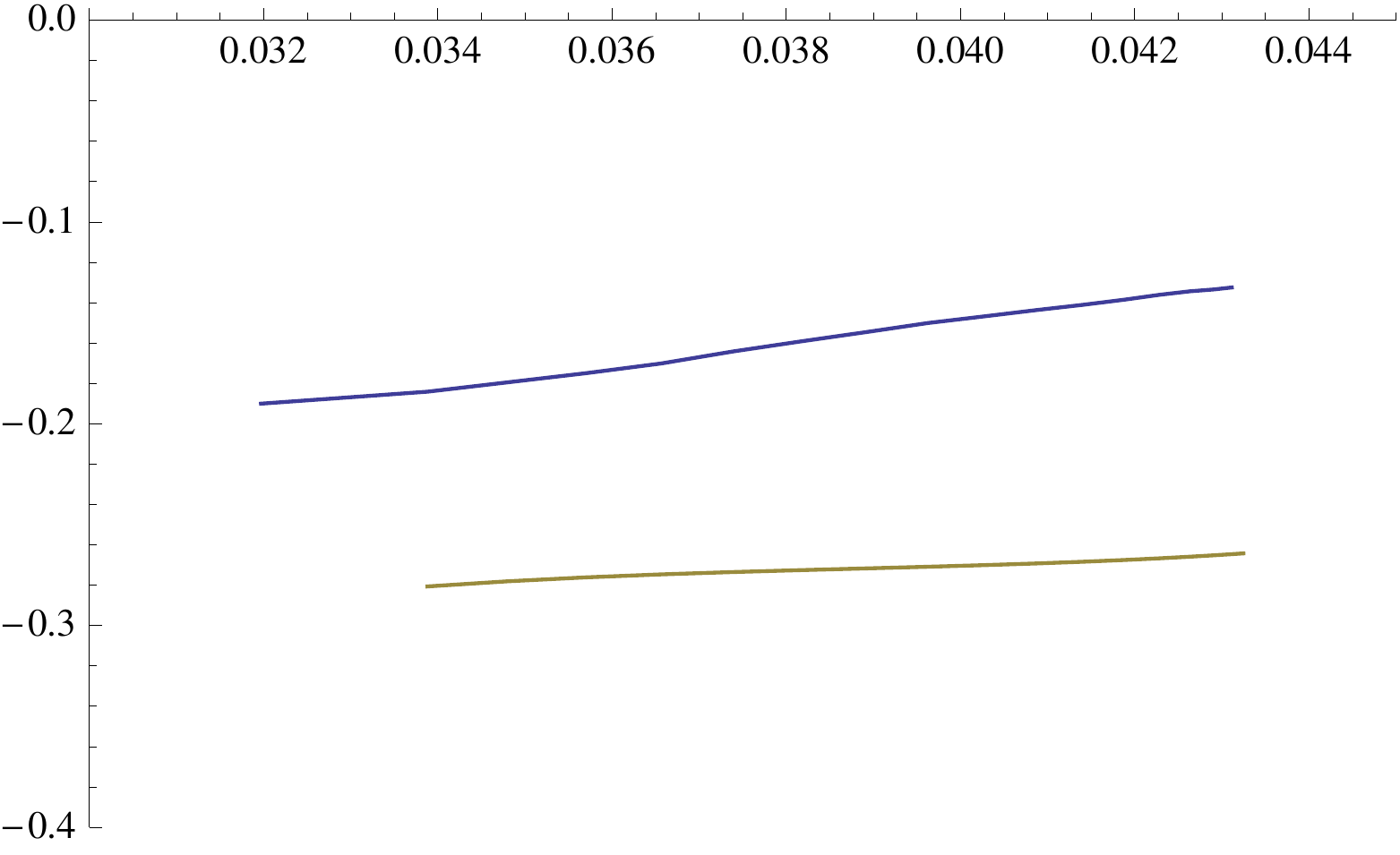}
\hskip2em
\includegraphics[width=0.4\textwidth]{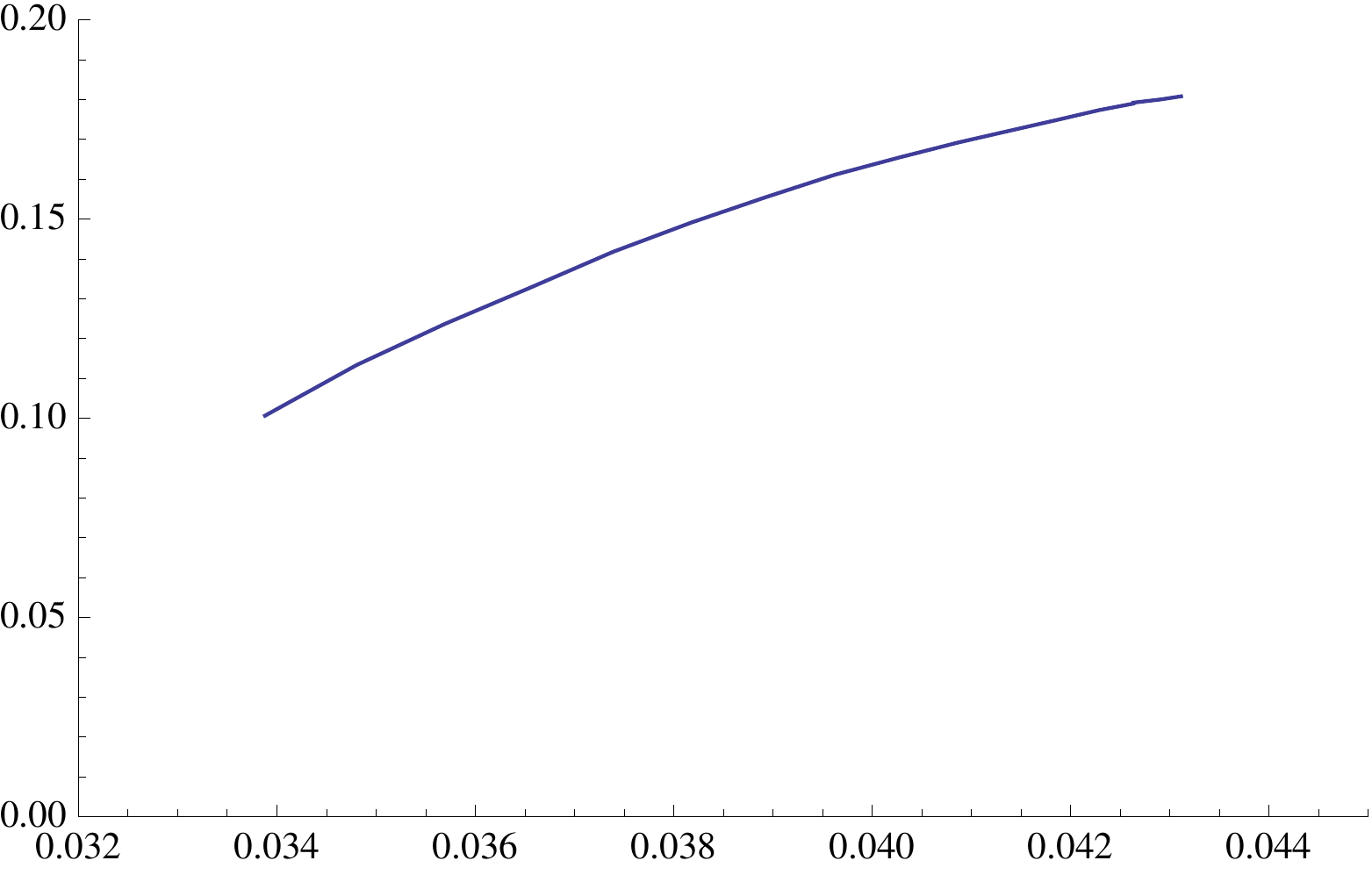}
\setlength{\unitlength}{0.1\textwidth}
\begin{picture}(0.3,0.4)(0,0)
\put(-9.1,2.){\makebox(0,0){$\Im\frac{\omega}{\mu}$}}
\put(-4.4,2.){\makebox(0,0){$d_{1,2}$}}
\put(0.2,0.2){\makebox(0,0){$\frac{T}{\mu}$}}
\end{picture}
\end{center}
\vskip-1em
\caption{Poles  in $G^R_{J^x J^x}$ closest to the real axis as a function of temperature for the broken insulator.  Left: We plot the purely-imaginary pole in blue and the (imaginary part of the) propagating pole in yellow. Right: The absolute value of the difference between these two poles, which is an order of magnitude smaller than for the broken metal --- see figure~\ref{fig:metal_QNM}.}\label{fig:insulator_QNM} 
\vskip-1em
\end{figure}

\section{Discussion}\label{sec:Discussion}

In this paper we studied the condensation of a charged scalar operator in two holographic models that allow momentum to relax. We  constructed  black branes   with charged scalar hair, corresponding to a broken phase, that are thermodynamically preferred over those without.  We demonstrated that the optical conductivity in a broken phase could often be described by a two-fluid model with a normal Drude component and a zero-frequency superfluid pole.  In some cases, however, the normal component was non-Drude, which was attributed to the lack of an isolated purely-dissipative QNM. It would be useful to study more points in the phase diagram, particularly in the axion-dilaton model, in order to obtain a complete picture. 

There are many similarities between our results and those found in other models, including the two-fluid model behavior at small frequencies, 
the rapid growth of the relaxation time as we lower the temperature, the pseudo gap in the broken metallic phase at low temperatures and 
the missing spectral weight manifested in the FGT sum rule.
However, some differences are apparent.  For example, whilst $T_c/\mu$ in the axion model decreases initially, we found that it increases for large enough values of $\alpha/\mu$, thus extending the results of \cite{Koga:2014hwa}.    
In addition, we find that the relaxation time in the broken metallic phase is not monotonic with the temperature, as opposed to the result presented in \cite{Horowitz:2013jaa}. 
Also, unlike  the inhomogeneous cases of \cite{Horowitz:2012gs, Donos:2014yya} and \cite{Horowitz:2013jaa}, we did not observe any resonances in the conductivity. As explained in \cite{Donos:2014yya}, such resonances are due to coupling of the current perturbations 
to the sound mode in the stress tensor correlators. 
In our homogeneous setups, however, we consider spatially-independent perturbations that decouple from the sound mode, accounting for the absence of resonances in this case.  It would be interesting to study the optical conductivity at finite momentum.

Next we discuss the interpretation of $K_n$ and $K_s$ appearing in \eqref{2 fluid model}.  In \cite{Horowitz:2013jaa}, these parameters were identified with the density of the normal and superfluid components of the fluid, respectively.   With this in mind, it is interesting to ask whether the system is fully condensed at zero temperature.  Proceeding naively and identifying $K_s$ with $\rho_s/\mu$, where $\rho_s$ is indeed a density, in the axion model we find  that $\rho_s/\rho$ tends to a constant   that is less than unity at low temperatures.\footnote{Here $\rho$ is the total charge density extracted from the fall-off of the gauge field in \eqref{eq:axion_gauge_field}.} This suggests that a normal component remains at zero temperature.  Furthermore, we observe that a greater normal component  remains for larger $\alpha/\mu$.  

However, whilst this identification is tempting to make,  one must be careful that the  normalization of the superfluid density is correct.  Furthermore, the connection with the quantity  that appeared as the  superfluid density in the  model of a holographic superfluid with finite supercurrent density in \cite{Sonner:2010yx}  is  not clear.  We expect that one must compute the optical conductivity in that model in order to resolve this issue.  If the normalization and interpretation  can indeed be fixed, then the axion-dilaton models studied here may provide a laboratory with sufficiently many parameters available in order to test `Homes' law' --- a universal relation between the superfluid density at zero temperature and the DC conductivity in the normal phase at the critical temperature \cite{Homes:2004wv, PhysRevB.72.134517}. This relation was first considered in a holographic context by \cite{Erdmenger:2012ik}.   We leave these interesting open questions for further work.

We conclude with a discussion of the generalizations that could be considered in the future.  We have chosen two examples from a general family of bulk actions with neutral sector 
\begin{equation}
	I_{\textrm{N}} = \int d^4 x \sqrt{-g} \left[ R - (\partial \phi)^2  -V_{\phi} (\phi) - \frac{Z(\phi)}{4} F^2 - Y(\phi)\sum_{i=1}^2  (\partial \chi_i)^2   \right ] 
	+ \int  X(\phi)\, F\wedge F
\end{equation}
and charged sector 
\begin{equation}
	I_{\textrm{C}} = \int d^4x \sqrt{-g} \left[ -  | D \psi  |^2 - V_{\psi}(|\psi|)  \right ] 
\end{equation}
There is great freedom in this class of models.  It would be worthwhile to work with a model that offers more control; for instance, with couplings that yield a simple IR geometry at zero temperature.  As a concrete example, it would be interesting to explore the superconducting phase transition in the model studied in \cite{Donos:2014uba} in which the normal phase at low temperatures is governed by a novel non-trivial fixed point in the IR.  In that model one could also study the effect on the broken phase when the normal phase is taken through a metal-insulator transition.  Moreover, alternative boundary conditions for the charged scalar could lead to additional effects, such as those described in \cite{Horowitz:2008bn}.
It would also be interesting to work directly with a more general model in order to understand which features are independent of the details of the model.

\section*{Acknowledgements}

It is a pleasure to thank Francesco Aprile, Richard Davison, Blaise Gout\'{e}raux, Elias Kiritsis, Kostas Skenderis, Marika Taylor and Benjamin Withers for useful discussions. 
The work of TA was supported by the European Research Council under the European Union's Seventh Framework Programme (ERC Grant agreement 307955).  
SAG was supported by National Science Foundation grant PHY-13-13986.
TA thanks the Physics Department at the University of Crete and the Mathematical Sciences Department at the University of Southampton for  hospitality during the completion of this work.


\providecommand{\href}[2]{#2}\begingroup\raggedright\endgroup

\end{document}